\documentclass{article}

\usepackage[preprint]{neurips_2025}

\usepackage[utf8]{inputenc} 
\usepackage[T1]{fontenc}    
\usepackage{hyperref}       
\usepackage{url}            
\usepackage{booktabs}       
\usepackage{amsfonts}       
\usepackage{nicefrac}       
\usepackage{microtype}      
\usepackage[table]{xcolor}         
\usepackage{graphicx} 
\usepackage[most]{tcolorbox}
\usepackage{enumitem}
\usepackage{subfig}

\title{PediatricsMQA: a Multi-modal Pediatrics Question Answering Benchmark}

\author{%
  Adil Bahaj\\
  Mohammed 6 Polytechnic University\\
  \texttt{adil.bahaj@um6p.ma} \\
  \And
   Oumaima Fadi\\
   International University of Rabat\\
   \texttt{oumaima.fadi@uir.ac.ma}
   \And
   Mohamed Chetouani\\
   Institut des Systèmes Intelligents et de Robotique \\
   \And
  Mounir Ghogho\\
  Mohammed 6 Polytechnic University\\
  \texttt{mounir.ghogho@um6p.ma} \\
}

\begin{document}

\maketitle

\begin{abstract}
  Large language models (LLMs) and vision-augmented LLMs (VLMs) have significantly advanced medical informatics, diagnostics, and decision support. However, these models exhibit systematic biases, particularly age bias, compromising their reliability and equity. This is evident in their poorer performance on pediatric-focused text and visual question-answering tasks. This bias reflects a broader imbalance in medical research, where pediatric studies receive less funding and representation despite the significant disease burden in children. To address these issues, a new comprehensive multi-modal pediatric question-answering benchmark, PediatricsMQA, has been introduced. It consists of 3,417 text-based multiple-choice questions (MCQs) covering 131 pediatric topics across seven developmental stages (prenatal to adolescent) and 2,067 vision-based MCQs using 634 pediatric images from 67 imaging modalities and 256 anatomical regions. The dataset was developed using a hybrid manual-automatic pipeline, incorporating peer-reviewed pediatric literature, validated question banks, existing benchmarks, and existing QA resources. Evaluating state‐of‐the‐art open models, we find dramatic performance drops in younger cohorts, highlighting the need for age‐aware methods to ensure equitable AI support in pediatric care.
\end{abstract}

\section{Introduction}
    Artificial intelligence (AI) has seen a rapid and widespread increase in its application to medical domains \cite{bahaj2025step, bahaj2024asthmabot, adil2022covid, elunveiling, bahajchildren}. These applications range from simple information management \cite{adil2022covid}, to detection and diagnostics \cite{elunveiling, alizadehsani2019machine} by using various techniques. Large language models (LLMs) and vision large language models (VLMs) are such techniques \cite{zhou2023survey}. LLMs and VLMs have been used extensively in specialised and generalist capacities in various medical applications \cite{zhou2023survey, thirunavukarasu2023large, kalpelbe2025vision}. However, multiple works found that these models suffer from different pitfalls (e.g. Hallucinations \cite{agarwal2024medhalu}, bias \cite{bahajchildren}, uncertainty \cite{xia2024cares}, privacy issues \cite{wiest2024privacy}, etc) that reduce their reliability. One of the most prominent issues is age bias. Age bias has been shown to exist at various levels of LLMs. In fact, \cite{bahajchildren} showed a significant age bias in known medical text question answering (TQA) benchmarks, which contain information more relevant to older than younger people. In addition, \cite{xia2024cares} showed that various VLMs perform in older age groups better than in younger age groups in the task of vision question answering (VQA). These works prove systemic neglect exists in LLM medical applications for younger people.

    This phenomenon is not exclusively present in LLMs, but it is rather an inherent part of medical research. In fact, various works \cite{martinez2008child, korinthenberg2023necessity} demonstrated that pediatrics is relatively understudied and under-explored. \cite{bourgeois2012pediatric} found that despite children accounting for nearly 60\% of the disease burden in certain conditions, only about 12\% of clinical drug trials focus on pediatric patients. \cite{martinez2008child} Found that studies involving adults are significantly more likely than child studies to be randomized controlled trials, systematic reviews, or studies of therapies. \cite{speer2023state} showed that pediatric research funding has historically been lower compared to adult research. These observed attitudes toward pediatric care inadvertently affect the quantity and quality of pediatric literature, which can influence the performance of LLMs in pediatrics. This neglect of child medicine can have significant ramifications on the adaptation of LLM-based applications in the case of pediatric care.
    
    In a step towards a fairer, more robust evaluation of LLMs and VLMs in pediatrics, this work proposes PediatricsMQA. PediatricsMQA contains TQA and VQA multi-choice questions (MCQs) and their corresponding right answers.  PediatricsMQA is constructed through a hybrid manual and automatic process from various sources (e.g. pediatrics journals articles, MCQ banks, pediatrics books, pediatrics QA pairs in previous benchmarks etc). PediatricsMQA is a rich representation of the field of pediatrics composed of TQA pairs 3417 about 131 pediatrics-related topics at 7 different child development stages (i.e. Prenate: -9 months to 0 weeks, Neonate: 0 weeks to 4 weeks, Infant: 1 month to 1 year, Toddler: 1–3 years, Preschool: 3–5 years, School-Age: 6–12 years, Adolescent: 13–18 years). In addition, PediatricsMQA contains 2067 VQA pairs on 634 images containing 67 modalities and 256 anatomical regions for male and female children at 7 different development stages. Extensive experiments on various publicly accessible LLMs show that PediatricsMQA represents a challenging benchmark. The dataset\footnote{\url{https://huggingface.co/datasets/adlbh/PediatricsMQA}} and code\footnote{\url{https://github.com/BahajAdil/PediatricsMQA}} are made available. Various findings were established as part of this work: \textbf{PediatricsMQA Difficulty}: The PediatricsMQA dataset is more challenging than other medical QA benchmarks, leading to lower model accuracies and revealing the complexity of pediatric reasoning. \textbf{Model Size Matters}: Newer and larger models (e.g., Llama-4-Maverick, Gemini-2.0-Flash) outperform smaller or older ones (e.g., MedAlpaca-7B), showing that model scale and sophistication are critical for pediatric QA. \textbf{Age Group Variability}: Model performance varies significantly across pediatric age groups. VQA models perform better on Neonates and Infants but struggle with Adolescents and Preschoolers. \textbf{Topic Sensitivity in TQA}: Models struggle with topics like “Lipid Disorders” and “Pharmacology,” but perform better on areas like “Developmental Psychology,” highlighting uneven reasoning across pediatric topics. \textbf{Anatomical Region Challenges in VQA}: Models excel with internal/frequently imaged regions (e.g., blood cells) but underperform on ambiguous or peripheral regions (e.g., gums, genitalia). \textbf{Modality Impact in VQA}: Structured modalities (e.g., optical images, IHC) yield higher accuracy, while complex ones (e.g., cytopathology, natural images) pose greater challenges. \textbf{Shared Limitations Across Models}: Regardless of architecture, all models exhibit similar weaknesses with certain modalities and topics, pointing to fundamental challenges in pediatric medical QA.
    
\section{Previous Literature}
    \subsection{LLMs and VLMs}
        The recent wave of advancement in natural language processing focused on instruction following LLMs \cite{ouyang2022training, alpaca, touvron2023llama}. These models showed remarkable performance and generality across a plethora of tasks (e.g. classification, summarisation, question answering, etc) and domains (e.g. finance, medicine, etc). Various LLMs were supplemented by a vision module to create VLMs. VLMs align vision and language to create an LLM with visual perception able to achieve various vision-related tasks (e.g. image captioning, VQA etc) in various domains (e.g. finance, medicine, law, etc). Since the inception of this new paradigm, various works have explored the feasibility and applicability of LLMs and VLMs to medical applications. Works such as \cite{xie2024me, wang2023huatuo, wu2024pmc} fine-tuned existing open-source LLMs on medical instruction data. \cite{chen2023meditron} pretrained and then finetuned an LLM on a large medical corpus. In addition, various works employed VLMs to solve medical vision tasks \cite{li2023llava, chen2024r, zhang2024development, hu2024omnimedvqa, gao2024mini, chen2024towards, alayrac2022flamingo}. Although these models demonstrated remarkable performance on different tasks, they were shown to suffer from various pitfalls. One of these pitfalls is age bias. In fact, \cite{bahajchildren} showed that TQA benchmark datasets contain information that is significantly more relevant to older people than younger people. \cite{xia2024cares} showed that VLMS demonstrated a significant age bias and performed relatively well on older people than younger people. This neglect of pediatric care motivated this work, which proposes a novel medical benchmark for pediatrics, containing TQA and VQA pairs.
        
    \subsection{Medical Benchmarks}
        Benchmark datasets are an important part of a model's life cycle, without which models cannot be evaluated on an equal footing. Various medical benchmarks have been proposed \cite{hu2024omnimedvqa} for various tasks (e.g. TQA \cite{pal2022medmcqa, jin2021disease, jin2019pubmedqa}, VQA \cite{hu2024omnimedvqa, zhang2024development}, summarisation \cite{deyoung2021msˆ2}, report generation \cite{yu2023evaluating, tian2023refisco}). QA benchmarks are particularly informative in medical applications because they quantify the factuality and exactitude of medical knowledge in LLMs and VLMs. However, works like \cite{bahajchildren} demonstrated the significant imbalance that exists between child and adult questions in various frequently used medical QA benchmarks. This limits the trustworthiness of existing LLMs and VLMs in pediatrics. This work tries to remedy this gap by proposing a new benchmark dataset for pediatric medicine. To our knowledge, there are only two datasets for pediatrics QA PediatricsQA \cite{bahajchildren} and PediaBench \cite{zhang2024pediabench}.  PediatricsMQA represents various advantages relative to these benchmarks. First, it extends TQA pairs of PediatricsQA and adds more TQA pairs in addition to containing highly curated VQA pairs. On the other hand, PediaBench only contains pediatrics TQA pairs in Chinese, which limits its adoption as a benchmark since most LLMs are trained on an English corpus.
    
\section{PediatricsMQA Dataset}
    PediatricsMQA contains: a) TQA and b) VQA pairs. These two types of questions in PediatricsMQA differ in their modality and the way they are constructed. Although there are various types of QA pairs in literature (e.g. binary QA pairs, MCQs pairs, open-ended QA pairs), PediatricsMQA contains only MCQs since they are harder than binary QA pairs, as they contain multiple options presenting a more challenging benchmark. In addition, MCQs are a more exact tool to assess the knowledge of LLMs than open-ended QA pairs. In what follows, we detail their construction process.
    \subsection{Construction process}
    \label{sec:const}
    \subsubsection{TQA Data Construction}
    This part of PediatricsMQA is built upon and extends PediatricsQA \cite{bahajchildren}. PediatricsQA contains 831 TQA pairs, which were extended in PediatricsMQA to 3401 TQA pairs. To construct the new TQA pairs, many medical sources were used (e.g. Books, question banks, medical exams, etc). We used various books that either contain QA pairs \cite{inbook} or that contain pediatrics medical knowledge \cite{behrman1983nelson}. TQA pairs are either made or extracted manually from books containing TQA pairs and pediatrics medical books, respectively. These questions belong to a diverse set of categories and child development stages, presenting various physical and psychological phenomena in pediatrics (Figures \ref{fig:category_count} and \ref{fig:age_group_count}). To avoid any unintentional copyright issues, we used an LLM (Gemini-2.0-Flash) to paraphrase the questions and the options (see appendix \ref{app:prompts}), in addition to changing the order of the options and the index of the right answer. Another round of manual filtering is conducted to remove any questions with obvious answers or questions that require contextual information to be answered. Although the vast majority of questions are about children, some questions about women's pregnancy are left due to their importance for prenate children's health.
    
    \subsubsection{VQA Data Construction}
    VQA pairs are extracted from two kinds of sources: a) based on previous benchmarks with demographic information (i.e. age), b) based on images and text from pediatrics journals.
    \paragraph{Based on existing benchmarks} Two benchmark datasets containing children's images are used: HAM10000 \cite{tschandl2018ham10000} and FairVLMed \cite{luo2024fairclip}. In both datasets, a subset of imaging and related metadata for people whose age is below 18 is taken. Since HAM10000 contains metadata about the localisation and diagnosis of the lesions present in the image, this metadata is used to create multiple VQA pairs through a template for the questions and options  (Figures \ref{fig:ham_gen_prompt_diag} and \ref{fig:ham_gen_prompt_loc}). The FairVLMed dataset medical notes that describe the image and the case, which are transformed to multi-choice questions using Gemini-2.0-Flash, by supplementing the LLM with the medical notes and the image and prompting it to produce 5 questions with their corresponding options and answers (Figure \ref{fig:fairvl_gen_prompt}). These MCQs are later filtered manually to leave only questions that can be answered visually and where the information is contained in the given clinical notes.
    
    \paragraph{Based on pediatrics journals} In this case, VQA pairs are extracted through the following steps: a) scraping, 2) generation, 3) curation. First, a set of images and their corresponding captions and passages from different scientific articles are obtained by scraping the journal of "Case Reports in Pediatrics" \cite{pedijournal}. This journal contains a plethora of articles about various cases of children with a diverse set of physical and psychological phenomena. Second, each image and its corresponding caption and passage are fed to a VLM (Gemini-2.0-Flash in our case) with a prompt instructing it to generate five relevant MCQs with their corresponding answer from the image in the context of the caption and the passage. Third, we noticed that different questions and answers were either out of context or were visually unanswerable (i.e. they require access to the caption and the passage to be answered). Consequently, manual curation was used to filter out low-quality QA pairs. Human labellers are tasked with adding two fields to each VQA pair: a) \textbf{question source} and b) \textbf{answer source}. The question source indicates if the question explicitly states that it requires access to the caption or passage, e.g. "based on the caption, ...". The answer source field indicates whether the answer is present in the caption and passages or is based on the internal knowledge of the LLM. The kept VQA pairs are those where the question is visual and the answer is contained in the context (i.e. passage + caption). An additional manual labeling step is conducted to extract the age and gender of each image from the article. All adult images are filtered out except those where a child is included (e.g. pregnant woman's CT scan or ultrasound).
    

    \subsection{Variations in the Data}
\begin{figure}
        \centering
        \includegraphics[scale=0.35]{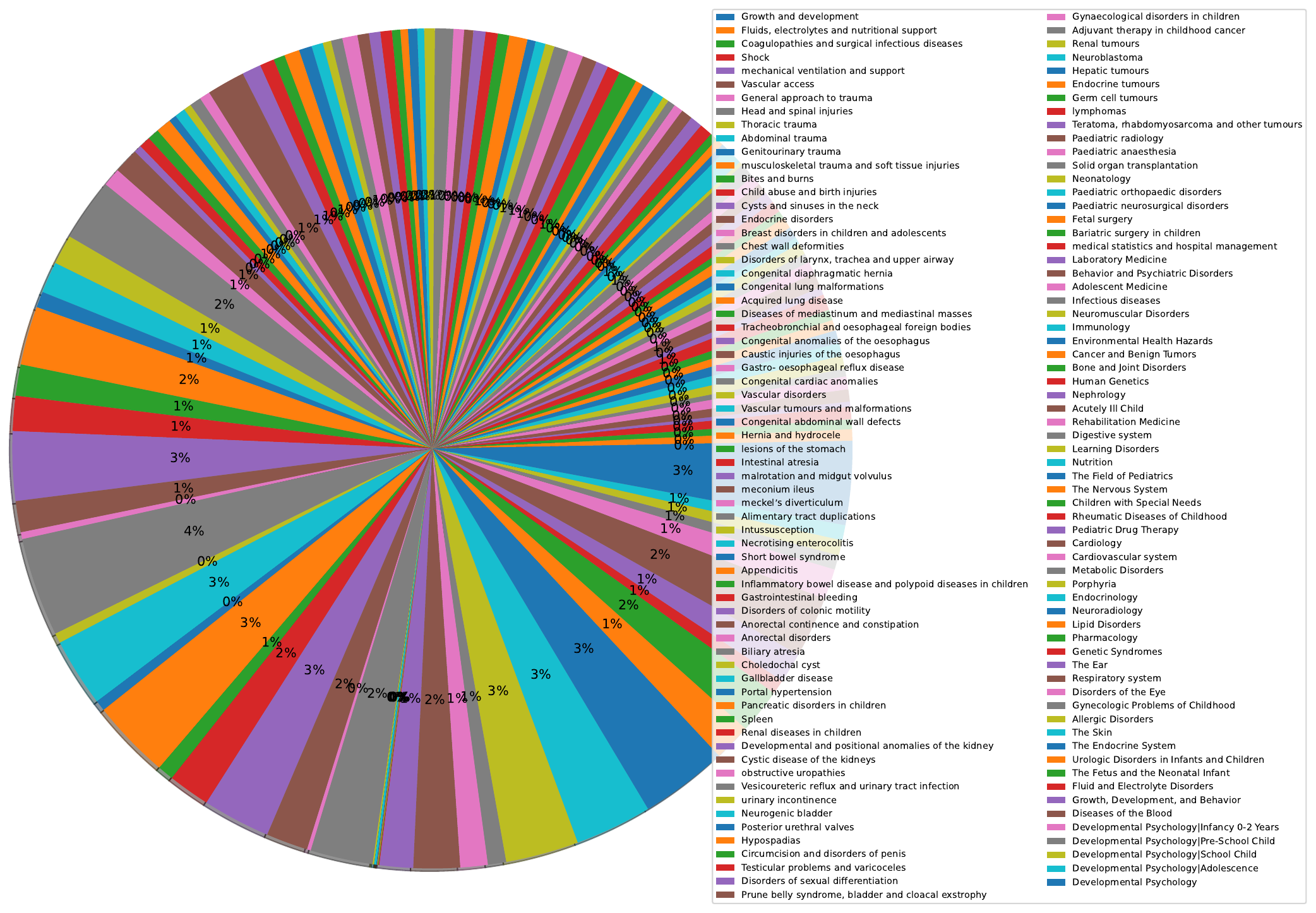}
        \caption{The different categories in TQA pairs in PediatricsMQA.}
        \label{fig:category_count}
    \end{figure}

\begin{figure}
        \centering
        \includegraphics[scale = 0.5]{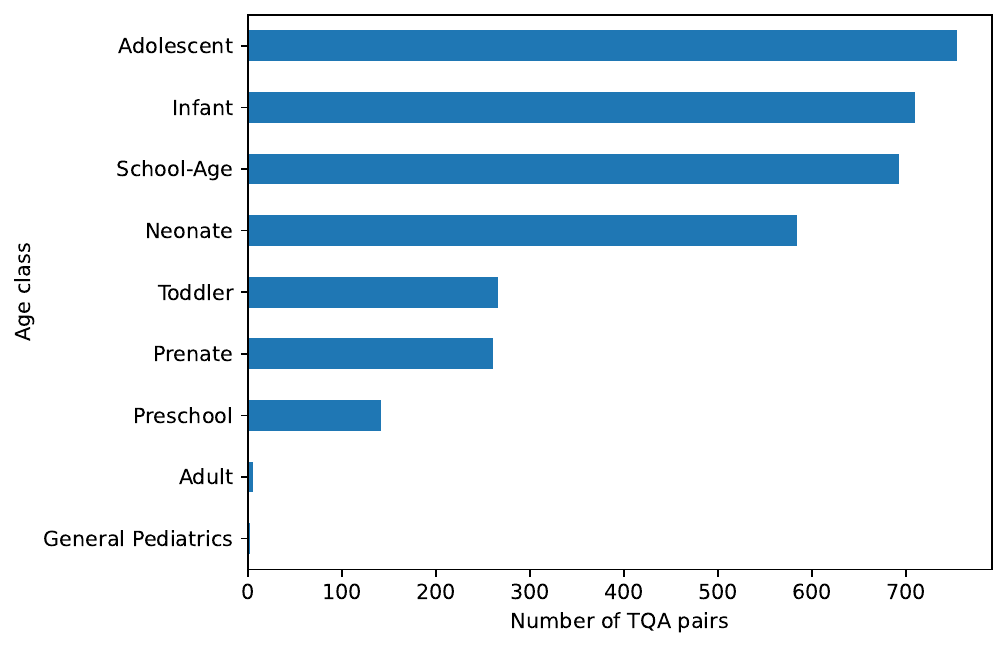}
        \caption{Distribution of different young people age groups in the TQA subset of PediatricsMQA. The following are the age group and their corresponding age range: Prenate: -9 months to 0 weeks, Neonate: 0 weeks to 4 weeks, Infant: 1 month to 1 year, Toddler: 1–3 years, Preschool: 3–5 years, School-Age: 6–12 years, Adolescent: 13–18 years}
        \label{fig:age_group_count}
    \end{figure}

\begin{figure}
        \centering
        \includegraphics[scale=0.5]{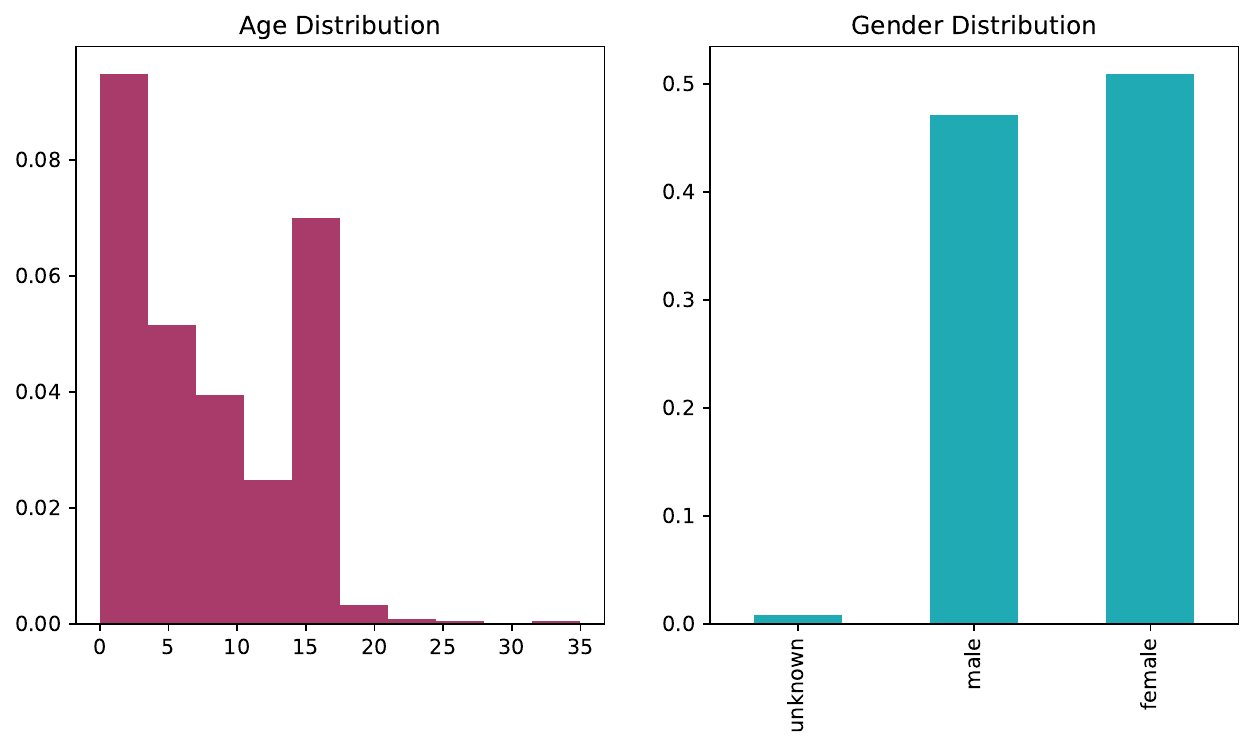}
        \caption{The distribution of age and gender in the TQA part of PediatricsMQA. The existence of some adult images and unknown genders is due to images of prenate babies which are still in gestation.}
        \label{fig:agegenderdist}
    \end{figure}

        The dataset contains various dimensions of variation. The following lists a few:
        \begin{itemize}
            \item \textbf{QA Tasks}: PediatricsMQA contains two subsets of TQA and VQA pairs, respectively. Each question is paired with multiple options (from 4 to 6 options), one of which is the right answer.
            \item \textbf{Categories and topics}: PediatricsMQA TQA pairs dataset contains a plethora of categories that deal with various topics ranging from basic psychological QA pairs to more advanced surgical QA pairs (Figure \ref{fig:category_count}).
            \item \textbf{Modalities}: PediatricsMQA VQA pairs belong to a diverse set of categories ranging from simple clinical imaging to more specialised imaging techniques like MRI and ultrasound (Table \ref{tab:ped_sum_stats}).
            \item \textbf{Demographics}: The dataset is also diverse across various demographic dimensions, including age and gender (Figures \ref{fig:agegenderdist}, \ref{fig:age_group_count}). In addition, VQA pairs of PediatricsMQA are about a range of internal and external anatomical regions (Table \ref{tab:ped_sum_stats}).
            \item \textbf{Medical Tasks}: PediatricsMQA QA pairs pertain to a range of medical tasks including: general knowledge, disease diagnosis, disease detection, anomaly detection etc
        \end{itemize}

        Table \ref{tab:ped_sum_stats} summarizes key stastics of the dataset. Appendix \ref{sec:dataset_app} shows a few examples from the dataset. 
        
    \begin{table}[]
        \centering
        \begin{tabular}{|c|c|c|c|c|c|}
            \hline
             \multicolumn{6}{|c|}{\cellcolor{blue!33}PediatricsMQA}  \\
             \hline
             \multicolumn{2}{|c}{\cellcolor{brown!33}TQA} & \multicolumn{4}{|c|}{\cellcolor{yellow!33}VQA}\\
             \hline
             \#QA pairs & \# Categories & \#QA pairs & \#Images &\# Anatomical regions &\# Modalities\\
             \hline
             3417&131&2067&634&256&67\\
             \hline
        \end{tabular}
        \caption{Summary Statistics. PediatricsMQA is composed of two main subsets of text QA pairs and vision QA pairs.}
        \label{tab:ped_sum_stats}
    \end{table}

\section{Results}
    \subsection{Benchmarks}
    Various flagship medical benchmarks are used to compare their performance to that of PediatricsMQA in TQA and VQA. For TQA, we report the performance of PediatricsQA \cite{bahajchildren}, PubMedQA \cite{jin2019pubmedqa}, MedQA \cite{jin2021disease}, and MedMCQA \cite{pal2022medmcqa}. For VQA, we report the performance of VQA-RAD \cite{lau2018dataset}, SLAKE \cite{liu2021slake} and PathVQA \cite{he2020pathvqa}.
    
    \subsection{Baselines}
    \label{sec:baselines}
    Different open and closed weight models are evaluated on TQA and VQA tasks for PediatricsMQA and other medical benchmarks. For TQA, we report the performance of MedAlpaca, Llama-Medx\footnote{\url{https://huggingface.co/skumar9/Llama-medx_v3.2}}, Gemini-1.5-Flash\cite{yang2024advancing, team2023gemini}, Gemini-2.0-Flash\cite{yang2024advancing, team2023gemini}, LLama-3.1 (8B) \cite{grattafiori2024llama}, Llama-4-scout (17B) \cite{metaLlamaHerd}, Llama-4-Maverick (17B) \cite{metaLlamaHerd}. For VQA, we report the performance of LLaVa-Med-7B \cite{li2023llava},
    HuatuoGPT-Vision-7B \cite{chen2024towards},
    Gemini-1.5-Flash\cite{yang2024advancing, team2023gemini}, and
    Gemini-2.0-Flash\cite{yang2024advancing, team2023gemini}.
    
    \subsection{Experimental setting}
    \label{sec:exp_set}
    The models are evaluated using the different benchmarks in addition to PediatricsMQA. We report the accuracy of each model in answering the different questions. For the VQA benchmarks, we only report the results of the closed setting, since it is a more comparable setting than the open setting. We use the default splits provided in the original papers corresponding to each dataset. All the datasets are evaluated using the test split, except MedMCQA, where we used the validation split since the test split doesn't have any ground truth labels. The experiments were conducted using various mediums. Some experiments were conducted in a local machine equipped with an RTX 3090 GPU. These experiments are in the following models: MedAlpaca, Llama-Medx, LLaVa-Med (7B), HuatuoGPT-Vision (7B). Other experiments were conducted using API calls for the Gemini models using Google AI Studio and Groq for the following models: LLama-3.1 (8B), Llama-4-scout (17B) and Llama-4-Maverick (17B).

    \subsection{Performance}
    Tables \ref{tab:tqa_results} and \ref{tab:vqa_results} compare the performance of various language and vision-language models, respectively. In the TQA results (Table \ref{tab:tqa_results}), the PediatricsMQA dataset proves to be a more challenging benchmark than established datasets like PubMedQA or MedQA, with overall lower scores across models. However, high-performing models such as Llama-4-Maverick and Gemini-2.0-Flash demonstrate strong capabilities, achieving top scores across nearly all datasets, including PediatricsMQA. In contrast, earlier or smaller-scale models like MedAlpaca1 and LLama-3.1 show significantly lower performance, especially on pediatric data, highlighting the added complexity of this domain. Similarly, in the VQA setting (Table \ref{fig:vqa_prompt}), PediatricsMQA again emerges as a more demanding testbed, particularly for models such as LLaVa-Med, which shows its weakest performance on this dataset. Nevertheless, Gemini-2.0-Flash and HuatuoGPT-Vision achieve competitive results, demonstrating their strength in handling multimodal medical content. Overall, the results underscore the relevance and difficulty of the PediatricsMQA benchmark and the increasing effectiveness of newer, more advanced multimodal models.
    \subsubsection{PediatricsMQA(TQA)}
    \paragraph{The effects of age on the performance}
        Table \ref{tab:age_group_tqa} illustrates the performance of various large language models on a pediatrics question answering dataset, with results broken down by different pediatric age groups: Prenate, Neonate, Infant, Toddler, Preschool, School-Age, and Adolescent. A consistent trend across all models is that performance varies between age groups, with no single age group being universally easier or harder for all models. Generally, the Llama-4-maverick (17b) and Gemini-2.0-flash models demonstrate the highest performance across most age categories, often achieving scores above 70\%. Conversely, MedAlpaca (7b) and Llama-3.1 (8b) consistently show lower performance, typically below 45\% across all age groups. For most models, there isn't a clear linear trend of increasing or decreasing performance with age; for instance, Llama-medx performs well in the Prenate and Adolescent groups but sees a dip in the Toddler category. Similarly, Gemini-1.5-flash shows relatively stable performance across most age groups with slight variations. The larger models, Llama-4-scout (17b) and Llama-4-maverick (17b), along with the Gemini series, generally outperform the smaller models like MedAlpaca (7b) and Llama-3.1 (8b), indicating a correlation between model size/sophistication and performance on this specialised medical question-answering task.   
    \paragraph{Effects of topic category on TQA performance}
        Tables \ref{tab:tqa_wb_cats_gemini_2.0}, \ref{tab:tqa_wb_cats_gemini_1.5}, \ref{tab:tqa_wb_cats_medalpaca-7b}, \ref{tab:tqa_wb_cats_llama-medx}, \ref{tab:tqa_wb_cats_llama_3.1-8b}, \ref{tab:tqa_wb_cats_llama-4-scout-17b}, and \ref{tab:tqa_wb_cats_llama-4-maverick-17b} represent a list of 10 worst and best performing PediatricqMQA(TQA) categories for Gemini-2.0-Flash, Gemini-1.0-Flash, MedAlpaca-7b, Llama-MedX, Llama-3.1 (8b), Llama-4-scout (17b) and Llama-4-maverick (17b) respectively. The results show that these models can score as low as 0\% in some categories, while performing as high as 100\% in other categories. In addition, models perform consistently badly at some categories, like "Lipid Disorders" and "Anorectal disorders", while being consistently good at other categories, such as "Developmental Psychology". This shows that there are some medical topics that various LLMs find consistently hard, requiring more training or better reasoning in these areas. These findings highlight both the potential and current limitations of LLMs in specialised pediatric medical reasoning, with implications for targeted dataset enrichment and model fine-tuning. The observed performance disparities across pediatric subdomains likely stem from variations in the complexity, specificity, and data representation of the topics within the models’ training corpora. Categories such as Lipid Disorders, Pharmacology, and Neuroradiology, which consistently appear among the worst-performing, often involve dense biochemical knowledge, imaging interpretation, or nuanced pharmacokinetics—areas that require detailed, structured understanding and are less likely to be covered extensively or uniformly in publicly available medical text data. Conversely, high-performing categories like Developmental Psychology, Porphyria, and Paediatric Orthopaedic Disorders may benefit from more narrative-style content, clearer clinical patterns, or broader representation in educational materials and online resources. Furthermore, the consistent success in Developmental Psychology suggests that topics grounded in developmental milestones and behavioral patterns—often discussed in accessible formats—are more easily internalised by language models. These trends indicate that performance may be correlated not only with domain difficulty but also with the nature and quality of training data available for each category. Targeted augmentation with high-quality domain-specific data could help address these disparities and improve model reliability across underperforming clinical areas.
    \subsubsection{PediatricsMQA(VQA)}
    \paragraph{Effects of age on the performance}
    Table \ref{tab:vqa_age_group_res} presents the performance of various vision-language models on the PediatricsMQA (Visual Question Answering) dataset, disaggregated by pediatric age groups: Prenate, Neonate, Infant, Toddler, Preschool, School-Age, and Adolescent. Among the models evaluated, Huatuogpt-vision (7B) consistently achieves the highest accuracy across most age groups, notably peaking at 73.77\% in the Neonate and 73.48\% in the Toddler categories. Gemini-2.0-flash also demonstrates relatively strong and stable performance, with scores ranging from 47.24\% to 68.18\%, outperforming its predecessor, Gemini-1.5-flash, across all age groups. In contrast, Llava-med-v1.5-mistral (7B) shows the lowest overall accuracy and more variable results, with its best performance (50.45\%) in the Infant category and a marked decline in Preschool (26.76\%). A general trend indicates that model performance tends to vary notably by age group, with the Neonate and Infant groups typically yielding higher accuracies, while performance tends to drop in the Adolescent and Preschool categories, highlighting the challenges models face in generalising across the developmental spectrum.
    \paragraph{Effects of anatomical region}

    The tables \ref{tab:vqa_bw_ana_gemini2.0}, \ref{tab:vqa_bw_ana_gemini1.5}, \ref{tab:vqa_bw_ana_huatuogpt_7b}, and \ref{tab:vqa_bw_ana_llava_v1.5_7b} present the best and worst performing anatomical regions in the PediatricsMQA(VQA) dataset across four different vision-language models: Gemini-2.0-flash, Gemini-1.5-flash, Huatuogpt-vision (7B), and Llava-med-v1.5-mistral (7B), respectively. Each table lists anatomical regions sorted by their question answering accuracy, with many regions consistently achieving perfect (100\%) or zero (0) accuracy. A cross-model comparison reveals several persistent trends. Notably, regions associated with cardiovascular and lymphatic structures, such as the coronary artery, vertebral body, blood cells, and buccal mucosa cells—frequently appear among the best performing across all models. Conversely, anatomical sites such as the gums, genital regions, parietal and occipital regions, axilla, and bronchus consistently fall among the worst-performing. These patterns suggest that models are more proficient in recognising and reasoning over medically salient or frequently imaged internal structures, whereas more peripheral, less frequently annotated, or visually ambiguous areas tend to challenge model performance uniformly.
    \paragraph{Effects of modality of the performance}

    The tables \ref{tab:vqa_bw_mod_gemini2.0}, \ref{tab:vqa_bw_mod_gemini1.5}, \ref{tab:vqa_bw_mod_huatuogpt7b}, and \ref{tab:vqa_bw_mod_llavav1.5} summarize the performance of various image modalities on the PediatricsMQA(VQA) dataset across four vision-language models: Gemini-2.0-flash, Gemini-1.5-flash, Huatuogpt-vision (7B), and Llava-med-v1.5-mistral (7B), respectively. Each table lists the ten best- and worst-performing modalities based on accuracy. A consistent trend across all models is the stark contrast in performance between modalities associated with visually rich, structured, or clinically interpretable data (e.g., optical, physical exam, surgical specimen, ihc staining)—which frequently achieve 100\% accuracy—and those involving abstract, low-contrast, or complex image types (e.g., natural image, cytopathology, vce, sanger sequencing)—which often yield 0\% accuracy. Additionally, modalities like radiograph, angiogram, and echocardiography appear among both the best and worst categories depending on the model, suggesting that model-specific strengths and weaknesses influence modality interpretation. This disparity highlights significant variability in how different vision-language architectures process and reason over medical imagery.
    \begin{table}[]
        \centering
        \setlength{\tabcolsep}{1.7pt}
        \begin{tabular}{l|p{6em}|p{5em}|p{4em}|p{3em}|p{5em}}
             Model & PediatricsMQA (TQA) & PediatricsQA \cite{bahajchildren} & PubMedQA \cite{jin2019pubmedqa}& MedQA \cite{jin2021disease}& MedMCQA \cite{pal2022medmcqa}\\
             \hline
             MedAlpaca\footnote{results for PubMedQA, MedQA and MedMCQA are from \url{https://github.com/MAGIC-AI4Med/MMedLM/tree/main}}&\cellcolor{red!33}31.05&35.13&72.8&41.7&37.5\\
             \hline
             Llama-Medx\footnote{results for PubMedQA, MedQA and MedMCQA are from \url{https://huggingface.co/spaces/openlifescienceai/open_medical_llm_leaderboard}.}&\cellcolor{red!33}50.43&55.18&73&61.04&60.53\\
             \hline
             Gemini-1.5-Flash\cite{yang2024advancing, team2023gemini}&\cellcolor{red!33}56.93&62.45&80.05&59.29&61.56\\
             \hline
             Gemini-2.0-Flash\cite{yang2024advancing, team2023gemini}&\cellcolor{red!33}70.61&74.54&90.1&79.0&70.77\\
             \hline
             LLama-3.1 (8B) \cite{grattafiori2024llama}&\cellcolor{red!33}38.53&42.96&55.20&44.87&43.24\\
             \hline
             Llama-4-scout (17B) \cite{metaLlamaHerd}&\cellcolor{red!33}66.14&67.74&70.03&74.69&68.80\\
             \hline
             Llama-4-Maverick (17B) \cite{metaLlamaHerd} &\cellcolor{red!33}\textbf{72.81}&\textbf{74.60}&\textbf{91.0}&\textbf{79.80}&\textbf{74.58}\\
             \hline
        \end{tabular}
        \caption{Results of TQA on various models for different datasets. Bold numbers represent the best-performing models in each dataset. The red cells represent the worst-performing dataset in each model.}
        \label{tab:tqa_results}
    \end{table}
    
    \begin{table}[]
        \centering
        \setlength{\tabcolsep}{1.7pt}
        \begin{tabular}{l|p{6em}|p{6em}|p{5em}|p{5em}}
            Model & PediatricsMQA (VQA) & VQA-RAD \cite{lau2018dataset} & SLAKE \cite{liu2021slake}& PathVQA \cite{he2020pathvqa}\\
            \hline
            LLaVa-Med (7B) \cite{li2023llava}& \cellcolor{red!33}34.95&51.4&48.6&56.8\\
            HuatuoGPT-Vision (7B) \cite{chen2024towards}& \cellcolor{red!33}\textbf{56.70}&63.7&76.2&57.9\\
            Gemini-1.5-Flash\cite{yang2024advancing, team2023gemini}&\cellcolor{red!33}30.01&64.94&73.80&66.27\\
            Gemini-2.0-Flash\cite{yang2024advancing, team2023gemini}&\cellcolor{red!33}55.0&\textbf{68.33}&\textbf{80.79}&\textbf{75.26}\\
            \hline
        \end{tabular}
        \caption{Results of VQA on various models for different datasets. Bold numbers represent the best-performing models in each dataset. The red cells represent the worst-performing dataset in each model.}
        \label{tab:vqa_results}
    \end{table}

\section{Limitations and future work}
    \label{sec:limit}
    We believe that pediatricsVQA offers a formidable and challenging benchmark for medical applications in pediatrics. However, this work can be extended and enriched in various ways. In what follows, we list a few directions that we intend to follow:
    \begin{itemize}
        \item \textbf{Pediatrics leaderboard}: Although the LLM and VLM models used in this work are representative of the larger landscape of existing open and closed models, a more thorough evaluation of more models can be beneficial. We intend to create a leaderboard of LLMs on pediatrics, which will be updated with new models and more challenging pediatrics data.
        \item \textbf{Enriching the data}: Although PediatricsMQA has a comparable size to existing evaluation benchmarks, it can always be grown by adding more questions, topics, images, modalities and other dimensions of variability.
        \item \textbf{Adding more modalities}: In addition to text and images, we envision a more comprehensive and inclusive dataset containing videos and audio also. This can be significantly important for various conditions like autism.
        \item \textbf{Reasoning}: creating a more sophisticated set of QA pairs that require reasoning and multi-step thinking to be answered correctly.
        \item \textbf{Training}: The same processes used to build PediatricsMQA can be scaled to produce a larger dataset, which can be used to train a pediatrics specialist LLM/VLM.
    \end{itemize}
    
\section{Conclusion}
This work introduces PediatricsMQA, a comprehensive benchmark designed to evaluate LLMs and VLMs in pediatric medical question answering. Addressing a critical gap in current medical AI research, PediatricsMQA captures the complexity and breadth of pediatric medicine across age groups, topics, anatomical regions, and imaging modalities. Extensive experiments show that pediatric content presents unique challenges, with consistently lower model performance compared to more general medical datasets. Model performance is influenced not only by scale and architecture but also by the nature of medical content—highlighting persistent weaknesses in underrepresented clinical domains and complex imaging types. These findings reaffirm the systemic neglect of pediatric-specific knowledge in existing AI systems and underscore the need for targeted dataset enrichment, pediatric-specific fine-tuning, and equitable evaluation strategies. PediatricsMQA serves as a step toward more inclusive, robust, and clinically reliable AI in healthcare.

\section*{Broader Impact}
\label{sec:broad_impact}
PediatricsMQA represents a major advancement in addressing the systemic neglect of pediatric needs in medical AI. By offering a comprehensive, age-stratified benchmark across diverse topics and modalities, it tackles the age bias in current LLMs/VLMs, which are typically adult-focused—an important step given children's historical underrepresentation in both medical research and AI training data.
\begin{itemize}[leftmargin=0.3cm]
    \item \textbf{Positive Societal Impacts}: PediatricsMQA can drive the creation of more equitable and reliable LLMs/VLMs for pediatric care, enabling AI tools that support clinical decisions, education, and patient communication across all age groups. By revealing model weaknesses across developmental stages, anatomy, and imaging types, it also advances safer AI deployment, with the potential to reduce diagnostic errors and improve outcomes in pediatric healthcare.
    \item \textbf{Negative Societal Impacts}: While this work offers valuable contributions, it also presents key risks. Increased focus on pediatric data heightens privacy concerns, especially if future dataset expansions lack strict ethical standards. There’s also a danger of over-reliance on LLMs/VLMs in sensitive clinical settings, where errors could cause real harm. Additionally, optimizing models for PediatricsMQA may lead to overfitting, limiting their effectiveness in real-world pediatric care.
\end{itemize}
In conclusion, while PediatricsMQA offers a critical tool for advancing fairness and capability in pediatric medical AI, it also necessitates responsible usage, continued human oversight, and proactive safeguards to mitigate ethical and societal risks.



\bibliographystyle{plain} 
\bibliography{references} 

\begin{thebibliography}{10}

\bibitem{adil2022covid}
Bahaj Adil, Safae Lhazmir, Mounir Ghogho, and Houda Benbrahim.
\newblock Covid-19-related scientific literature exploration: Short survey and comparative study.
\newblock {\em Biology}, 11(8):1221, 2022.

\bibitem{agarwal2024medhalu}
Vibhor Agarwal, Yiqiao Jin, Mohit Chandra, Munmun De~Choudhury, Srijan Kumar, and Nishanth Sastry.
\newblock Medhalu: Hallucinations in responses to healthcare queries by large language models.
\newblock {\em arXiv preprint arXiv:2409.19492}, 2024.

\bibitem{alayrac2022flamingo}
Jean-Baptiste Alayrac, Jeff Donahue, Pauline Luc, Antoine Miech, Iain Barr, Yana Hasson, Karel Lenc, Arthur Mensch, Katherine Millican, Malcolm Reynolds, et~al.
\newblock Flamingo: a visual language model for few-shot learning.
\newblock {\em Advances in neural information processing systems}, 35:23716--23736, 2022.

\bibitem{alizadehsani2019machine}
Roohallah Alizadehsani, Moloud Abdar, Mohamad Roshanzamir, Abbas Khosravi, Parham~M Kebria, Fahime Khozeimeh, Saeid Nahavandi, Nizal Sarrafzadegan, and U~Rajendra Acharya.
\newblock Machine learning-based coronary artery disease diagnosis: A comprehensive review.
\newblock {\em Computers in biology and medicine}, 111:103346, 2019.

\bibitem{inbook}
Zuhair Almusawi and Hasanein Ghali.
\newblock {\em MCQ in Pediatrics - Review of Nelson textbook of Pediatrics}.
\newblock 10 2016.

\bibitem{bahajchildren}
Adil Bahaj, Mohamed CHETOUANI, and Mounir Ghogho.
\newblock What about the children? evaluating and mitigating ageism in medical qa benchmarks.
\newblock In {\em ICLR AI4CHL Workshop}, 2025.

\bibitem{bahaj2024asthmabot}
Adil Bahaj and Mounir Ghogho.
\newblock Asthmabot: Multi-modal, multi-lingual retrieval augmented generation for asthma patient support.
\newblock {\em arXiv preprint arXiv:2409.15815}, 2024.

\bibitem{bahaj2025step}
Adil Bahaj and Mounir Ghogho.
\newblock A step towards quantifying, modelling and exploring uncertainty in biomedical knowledge graphs.
\newblock {\em Computers in Biology and Medicine}, 184:109355, 2025.

\bibitem{behrman1983nelson}
Richard~E Behrman and III Vaughan, VC.
\newblock {\em Nelson textbook of pediatrics.}
\newblock Number Ed. 12. 1983.

\bibitem{bourgeois2012pediatric}
Florence~T Bourgeois, Srinivas Murthy, Catia Pinto, Karen~L Olson, John~PA Ioannidis, and Kenneth~D Mandl.
\newblock Pediatric versus adult drug trials for conditions with high pediatric disease burden.
\newblock {\em Pediatrics}, 130(2):285--292, 2012.

\bibitem{chen2024towards}
Junying Chen, Chi Gui, Ruyi Ouyang, Anningzhe Gao, Shunian Chen, Guiming Chen, Xidong Wang, Zhenyang Cai, Ke~Ji, Xiang Wan, et~al.
\newblock Towards injecting medical visual knowledge into multimodal llms at scale.
\newblock In {\em Proceedings of the 2024 Conference on Empirical Methods in Natural Language Processing}, pages 7346--7370, 2024.

\bibitem{chen2024r}
Xupeng Chen, Zhixin Lai, Kangrui Ruan, Shichu Chen, Jiaxiang Liu, and Zuozhu Liu.
\newblock R-llava: Improving med-vqa understanding through visual region of interest.
\newblock {\em arXiv preprint arXiv:2410.20327}, 2024.

\bibitem{chen2023meditron}
Zeming Chen, Alejandro~Hern{\'a}ndez Cano, Angelika Romanou, Antoine Bonnet, Kyle Matoba, Francesco Salvi, Matteo Pagliardini, Simin Fan, Andreas K{\"o}pf, Amirkeivan Mohtashami, et~al.
\newblock Meditron-70b: Scaling medical pretraining for large language models.
\newblock {\em arXiv preprint arXiv:2311.16079}, 2023.

\bibitem{deyoung2021msˆ2}
Jay DeYoung, Iz~Beltagy, Madeleine van Zuylen, Bailey Kuehl, and Lucy~Lu Wang.
\newblock Msˆ2: Multi-document summarization of medical studies.
\newblock In {\em Proceedings of the 2021 Conference on Empirical Methods in Natural Language Processing}, pages 7494--7513, 2021.

\bibitem{elunveiling}
Hasna El~Haji, Nada Sbihi, Kaoutar El~Handri, Adil Bahaj, Mohammed Elkasri, Amine Souadka, and Mounir Ghogho.
\newblock Unveiling breast cancer causes through knowledge graph analysis and biobert-based factuality prediction.
\newblock 2025.

\bibitem{gao2024mini}
Zhangwei Gao, Zhe Chen, Erfei Cui, Yiming Ren, Weiyun Wang, Jinguo Zhu, Hao Tian, Shenglong Ye, Junjun He, Xizhou Zhu, et~al.
\newblock Mini-internvl: a flexible-transfer pocket multi-modal model with 5\% parameters and 90\% performance.
\newblock {\em Visual Intelligence}, 2(1):1--17, 2024.

\bibitem{grattafiori2024llama}
Aaron Grattafiori, Abhimanyu Dubey, Abhinav Jauhri, Abhinav Pandey, Abhishek Kadian, Ahmad Al-Dahle, Aiesha Letman, Akhil Mathur, Alan Schelten, Alex Vaughan, et~al.
\newblock The llama 3 herd of models.
\newblock {\em arXiv preprint arXiv:2407.21783}, 2024.

\bibitem{he2020pathvqa}
Xuehai He, Yichen Zhang, Luntian Mou, Eric Xing, and Pengtao Xie.
\newblock Pathvqa: 30000+ questions for medical visual question answering.
\newblock {\em arXiv preprint arXiv:2003.10286}, 2020.

\bibitem{hu2024omnimedvqa}
Yutao Hu, Tianbin Li, Quanfeng Lu, Wenqi Shao, Junjun He, Yu~Qiao, and Ping Luo.
\newblock Omnimedvqa: A new large-scale comprehensive evaluation benchmark for medical lvlm.
\newblock In {\em Proceedings of the IEEE/CVF Conference on Computer Vision and Pattern Recognition}, pages 22170--22183, 2024.

\bibitem{jin2021disease}
Di~Jin, Eileen Pan, Nassim Oufattole, Wei-Hung Weng, Hanyi Fang, and Peter Szolovits.
\newblock What disease does this patient have? a large-scale open domain question answering dataset from medical exams.
\newblock {\em Applied Sciences}, 11(14):6421, 2021.

\bibitem{jin2019pubmedqa}
Qiao Jin, Bhuwan Dhingra, Zhengping Liu, William Cohen, and Xinghua Lu.
\newblock Pubmedqa: A dataset for biomedical research question answering.
\newblock In {\em Proceedings of the 2019 Conference on Empirical Methods in Natural Language Processing and the 9th International Joint Conference on Natural Language Processing (EMNLP-IJCNLP)}, pages 2567--2577, 2019.

\bibitem{pedijournal}
Journal.
\newblock Case reports in pediatrics.
\newblock \url{https://onlinelibrary.wiley.com/journal/2920}.
\newblock [Accessed 29-04-2025].

\bibitem{kalpelbe2025vision}
B{\'e}ria~Chingnab{\'e} Kalp{\'e}lb{\'e}, Angel~Gabriel Adaambiik, and Wei Peng.
\newblock Vision language models in medicine.
\newblock {\em arXiv preprint arXiv:2503.01863}, 2025.

\bibitem{korinthenberg2023necessity}
Rudolf Korinthenberg.
\newblock Necessity and limitations of paediatric research—a personal view.
\newblock {\em Pediatric Medicine}, 6, 2023.

\bibitem{lau2018dataset}
Jason~J Lau, Soumya Gayen, Asma Ben~Abacha, and Dina Demner-Fushman.
\newblock A dataset of clinically generated visual questions and answers about radiology images.
\newblock {\em Scientific data}, 5(1):1--10, 2018.

\bibitem{li2023llava}
Chunyuan Li, Cliff Wong, Sheng Zhang, Naoto Usuyama, Haotian Liu, Jianwei Yang, Tristan Naumann, Hoifung Poon, and Jianfeng Gao.
\newblock Llava-med: Training a large language-and-vision assistant for biomedicine in one day.
\newblock {\em Advances in Neural Information Processing Systems}, 36:28541--28564, 2023.

\bibitem{liu2021slake}
Bo~Liu, Li-Ming Zhan, Li~Xu, Lin Ma, Yan Yang, and Xiao-Ming Wu.
\newblock Slake: A semantically-labeled knowledge-enhanced dataset for medical visual question answering.
\newblock In {\em 2021 IEEE 18th international symposium on biomedical imaging (ISBI)}, pages 1650--1654. IEEE, 2021.

\bibitem{luo2024fairclip}
Yan Luo, Min Shi, Muhammad~Osama Khan, Muhammad~Muneeb Afzal, Hao Huang, Shuaihang Yuan, Yu~Tian, Luo Song, Ava Kouhana, Tobias Elze, et~al.
\newblock Fairclip: Harnessing fairness in vision-language learning.
\newblock In {\em Proceedings of the IEEE/CVF Conference on Computer Vision and Pattern Recognition}, pages 12289--12301, 2024.

\bibitem{martinez2008child}
Carolina Martinez-Castaldi, Michael Silverstein, and Howard Bauchner.
\newblock Child versus adult research: the gap in high-quality study design.
\newblock {\em Pediatrics}, 122(1):52--57, 2008.

\bibitem{ouyang2022training}
Long Ouyang, Jeffrey Wu, Xu~Jiang, Diogo Almeida, Carroll Wainwright, Pamela Mishkin, Chong Zhang, Sandhini Agarwal, Katarina Slama, Alex Ray, et~al.
\newblock Training language models to follow instructions with human feedback.
\newblock {\em Advances in neural information processing systems}, 35:27730--27744, 2022.

\bibitem{pal2022medmcqa}
Ankit Pal, Logesh~Kumar Umapathi, and Malaikannan Sankarasubbu.
\newblock Medmcqa: A large-scale multi-subject multi-choice dataset for medical domain question answering.
\newblock In {\em Conference on health, inference, and learning}, pages 248--260. PMLR, 2022.

\bibitem{speer2023state}
Esther~M Speer, Lois~K Lee, Florence~T Bourgeois, Daniel Gitterman, William~W Hay~Jr, Jonathan~M Davis, and Joyce~R Javier.
\newblock The state and future of pediatric research—an introductory overview: The state and future of pediatric research series.
\newblock {\em Pediatric Research}, pages 1--5, 2023.

\bibitem{alpaca}
Rohan Taori, Ishaan Gulrajani, Tianyi Zhang, Yann Dubois, Xuechen Li, Carlos Guestrin, Percy Liang, and Tatsunori~B. Hashimoto.
\newblock Stanford alpaca: An instruction-following llama model.
\newblock \url{https://github.com/tatsu-lab/stanford_alpaca}, 2023.

\bibitem{team2023gemini}
Gemini Team, Rohan Anil, Sebastian Borgeaud, Jean-Baptiste Alayrac, Jiahui Yu, Radu Soricut, Johan Schalkwyk, Andrew~M Dai, Anja Hauth, Katie Millican, et~al.
\newblock Gemini: a family of highly capable multimodal models.
\newblock {\em arXiv preprint arXiv:2312.11805}, 2023.

\bibitem{metaLlamaHerd}
Meta~Llama Team.
\newblock {T}he {L}lama 4 herd: {T}he beginning of a new era of natively multimodal {A}{I} innovation --- ai.meta.com.
\newblock \url{https://ai.meta.com/blog/llama-4-multimodal-intelligence/}, 2025.
\newblock [Accessed 05-05-2025].

\bibitem{thirunavukarasu2023large}
Arun~James Thirunavukarasu, Darren Shu~Jeng Ting, Kabilan Elangovan, Laura Gutierrez, Ting~Fang Tan, and Daniel Shu~Wei Ting.
\newblock Large language models in medicine.
\newblock {\em Nature medicine}, 29(8):1930--1940, 2023.

\bibitem{tian2023refisco}
Katherine Tian, Sina~J Hartung, Andrew~A Li, Jaehwan Jeong, Fardad Behzadi, Juan Calle-Toro, Subathra Adithan, Michael Pohlen, David Osayande, and Pranav Rajpurkar.
\newblock Refisco: Report fix and score dataset for radiology report generation, 2023.

\bibitem{touvron2023llama}
Hugo Touvron, Thibaut Lavril, Gautier Izacard, Xavier Martinet, Marie-Anne Lachaux, Timoth{\'e}e Lacroix, Baptiste Rozi{\`e}re, Naman Goyal, Eric Hambro, Faisal Azhar, et~al.
\newblock Llama: Open and efficient foundation language models.
\newblock {\em arXiv preprint arXiv:2302.13971}, 2023.

\bibitem{tschandl2018ham10000}
Philipp Tschandl, Cliff Rosendahl, and Harald Kittler.
\newblock The ham10000 dataset, a large collection of multi-source dermatoscopic images of common pigmented skin lesions.
\newblock {\em Scientific data}, 5(1):1--9, 2018.

\bibitem{wang2023huatuo}
Haochun Wang, Chi Liu, Nuwa Xi, Zewen Qiang, Sendong Zhao, Bing Qin, and Ting Liu.
\newblock Huatuo: Tuning llama model with chinese medical knowledge.
\newblock {\em arXiv preprint arXiv:2304.06975}, 2023.

\bibitem{wiest2024privacy}
Isabella~Catharina Wiest, Dyke Ferber, Jiefu Zhu, Marko van Treeck, Sonja~K Meyer, Radhika Juglan, Zunamys~I Carrero, Daniel Paech, Jens Kleesiek, Matthias~P Ebert, et~al.
\newblock Privacy-preserving large language models for structured medical information retrieval.
\newblock {\em NPJ Digital Medicine}, 7(1):257, 2024.

\bibitem{wu2024pmc}
Chaoyi Wu, Weixiong Lin, Xiaoman Zhang, Ya~Zhang, Weidi Xie, and Yanfeng Wang.
\newblock Pmc-llama: toward building open-source language models for medicine.
\newblock {\em Journal of the American Medical Informatics Association}, 31(9):1833--1843, 2024.

\bibitem{xia2024cares}
Peng Xia, Ze~Chen, Juanxi Tian, Yangrui Gong, Ruibo Hou, Yue Xu, Zhenbang Wu, Zhiyuan Fan, Yiyang Zhou, Kangyu Zhu, et~al.
\newblock Cares: A comprehensive benchmark of trustworthiness in medical vision language models.
\newblock {\em Advances in Neural Information Processing Systems}, 37:140334--140365, 2024.

\bibitem{xie2024me}
Qianqian Xie, Qingyu Chen, Aokun Chen, Cheng Peng, Yan Hu, Fongci Lin, Xueqing Peng, Jimin Huang, Jeffrey Zhang, Vipina Keloth, et~al.
\newblock Me-llama: Foundation large language models for medical applications.
\newblock {\em Research square}, pages rs--3, 2024.

\bibitem{yang2024advancing}
Lin Yang, Shawn Xu, Andrew Sellergren, Timo Kohlberger, Yuchen Zhou, Ira Ktena, Atilla Kiraly, Faruk Ahmed, Farhad Hormozdiari, Tiam Jaroensri, et~al.
\newblock Advancing multimodal medical capabilities of gemini.
\newblock {\em arXiv preprint arXiv:2405.03162}, 2024.

\bibitem{yu2023evaluating}
Feiyang Yu, Mark Endo, Rayan Krishnan, Ian Pan, Andy Tsai, Eduardo~Pontes Reis, Eduardo Kaiser Ururahy~Nunes Fonseca, Henrique Min~Ho Lee, Zahra Shakeri~Hossein Abad, Andrew~Y Ng, et~al.
\newblock Evaluating progress in automatic chest x-ray radiology report generation.
\newblock {\em Patterns}, 4(9), 2023.

\bibitem{zhang2024pediabench}
Qian Zhang, Panfeng Chen, Jiali Li, Linkun Feng, Shuyu Liu, Mei Chen, Hui Li, and Yanhao Wang.
\newblock Pediabench: A comprehensive chinese pediatric dataset for benchmarking large language models.
\newblock {\em arXiv preprint arXiv:2412.06287}, 2024.

\bibitem{zhang2024development}
Xiaoman Zhang, Chaoyi Wu, Ziheng Zhao, Weixiong Lin, Ya~Zhang, Yanfeng Wang, and Weidi Xie.
\newblock Development of a large-scale medical visual question-answering dataset.
\newblock {\em Communications Medicine}, 4(1):277, 2024.

\bibitem{zhou2023survey}
Hongjian Zhou, Fenglin Liu, Boyang Gu, Xinyu Zou, Jinfa Huang, Jinge Wu, Yiru Li, Sam~S Chen, Peilin Zhou, Junling Liu, et~al.
\newblock A survey of large language models in medicine: Progress, application, and challenge.
\newblock {\em arXiv preprint arXiv:2311.05112}, 2023.

\end{thebibliography}


\newpage
\section*{NeurIPS Paper Checklist}

\begin{enumerate}

\item {\bf Claims}
    \item[] Question: Do the main claims made in the abstract and introduction accurately reflect the paper's contributions and scope?
    \item[] Answer: \answerYes{} 
    \item[] Justification: The paper identifies a systemic pediatric underrepresentation problem in LLMs/VLMs and proposes PediatricsMQA as a new benchmark to evaluate and address this issue. Various LLM and VLM models are evaluated on PediatricsMQA and were shown to find this benchmark more challenging than other benchmarks of comparable sizes.

\item {\bf Limitations}
    \item[] Question: Does the paper discuss the limitations of the work performed by the authors?
    \item[] Answer: \answerYes{} 
    \item[] Justification: Section \ref{sec:limit} discusses different limitations that the paper has and potential future work to surpass them.
    

\item {\bf Theory assumptions and proofs}
    \item[] Question: For each theoretical result, does the paper provide the full set of assumptions and a complete (and correct) proof?
    \item[] Answer: \answerNA{} 
    \item[] Justification: This paper proposes a benchmark. We focused on empirical studies rather than any theoretical exploration.

    \item {\bf Experimental result reproducibility}
    \item[] Question: Does the paper fully disclose all the information needed to reproduce the main experimental results of the paper to the extent that it affects the main claims and/or conclusions of the paper (regardless of whether the code and data are provided or not)?
    \item[] Answer: \answerYes{} 
    \item[] Justification: The data construction process is described in detail in section \ref{sec:const}. In addition to giving the different prompts used to generate the data in section \ref{app:prompts}. The models used for evaluation are also disclosed in section \ref{sec:baselines}.

\item {\bf Open access to data and code}
    \item[] Question: Does the paper provide open access to the data and code, with sufficient instructions to faithfully reproduce the main experimental results, as described in supplemental material?
    \item[] Answer: \answerYes{} 
    \item[] Justification: The code and data are made available following the code and data submission guidelines of the "datasets and benchmarks" track.

\item {\bf Experimental setting/details}
    \item[] Question: Does the paper specify all the training and test details (e.g., data splits, hyperparameters, how they were chosen, type of optimizer, etc.) necessary to understand the results?
    \item[] Answer: \answerYes{} 
    \item[] Justification: we provide the experimental settings in section \ref{sec:exp_set}.

\item {\bf Experiment statistical significance}
    \item[] Question: Does the paper report error bars suitably and correctly defined or other appropriate information about the statistical significance of the experiments?
    \item[] Answer: \answerNo{} 
    \item[] Justification: We didn't report statistical significance results. However, due to the general consistency of LLMs we hypothesise that the results won't vary between runs, especially in flagship models like Gemini.

\item {\bf Experiments compute resources}
    \item[] Question: For each experiment, does the paper provide sufficient information on the computer resources (type of compute workers, memory, time of execution) needed to reproduce the experiments?
    \item[] Answer: \answerYes{} 
    \item[] Justification: Section \ref{sec:exp_set} provides that information.
    
\item {\bf Code of ethics}
    \item[] Question: Does the research conducted in the paper conform, in every respect, with the NeurIPS Code of Ethics \url{https://neurips.cc/public/EthicsGuidelines}?
    \item[] Answer: \answerYes{} 
    \item[] Justification: We carefully observe the code of ethics during our work.

\item {\bf Broader impacts}
    \item[] Question: Does the paper discuss both potential positive societal impacts and negative societal impacts of the work performed?
    \item[] Answer: \answerYes{} 
    \item[] Justification: Section \ref{sec:broad_impact} lists the broader positive and negative impacts of this work.
    
\item {\bf Safeguards}
    \item[] Question: Does the paper describe safeguards that have been put in place for responsible release of data or models that have a high risk for misuse (e.g., pretrained language models, image generators, or scraped datasets)?
    \item[] Answer: \answerNo{} 
    \item[] Justification: The dataset is scraped from previous public benchmarks and public open-access journals and sources. However, due to our reliance on LLMs to generate the data, we had to incorporate a manual curation process to remove any harmful or irrelevant questions and answers, which we described in section \ref{sec:const}.

\item {\bf Licenses for existing assets}
    \item[] Question: Are the creators or original owners of assets (e.g., code, data, models), used in the paper, properly credited and are the license and terms of use explicitly mentioned and properly respected?
    \item[] Answer: \answerYes{} 
    \item[] Justification: all original creates are credited along the paper.

\item {\bf New assets}
    \item[] Question: Are new assets introduced in the paper well documented and is the documentation provided alongside the assets?
    \item[] Answer: \answerYes{} 
    \item[] Justification: The dataset and code are documented.

\item {\bf Crowdsourcing and research with human subjects}
    \item[] Question: For crowdsourcing experiments and research with human subjects, does the paper include the full text of instructions given to participants and screenshots, if applicable, as well as details about compensation (if any)? 
    \item[] Answer: \answerNA{} 
    \item[] Justification: Doesn't apply to us since we didn't have any human subjects or crowdsourced our process.

\item {\bf Institutional review board (IRB) approvals or equivalent for research with human subjects}
    \item[] Question: Does the paper describe potential risks incurred by study participants, whether such risks were disclosed to the subjects, and whether Institutional Review Board (IRB) approvals (or an equivalent approval/review based on the requirements of your country or institution) were obtained?
    \item[] Answer: \answerNA{} 
    \item[] Justification: doesn't apply.

\item {\bf Declaration of LLM usage}
    \item[] Question: Does the paper describe the usage of LLMs if it is an important, original, or non-standard component of the core methods in this research? Note that if the LLM is used only for writing, editing, or formatting purposes and does not impact the core methodology, scientific rigorousness, or originality of the research, declaration is not required.
    \item[] Answer: \answerYes{} 
    \item[] Justification: We used LLMs in the data construction and evaluation process, which we described in section \ref{sec:const}.

\end{enumerate}

\newpage

\appendix

\section{Dataset}
    \label{sec:dataset_app}
\begin{figure}
        \centering
        \includegraphics[scale=0.4]{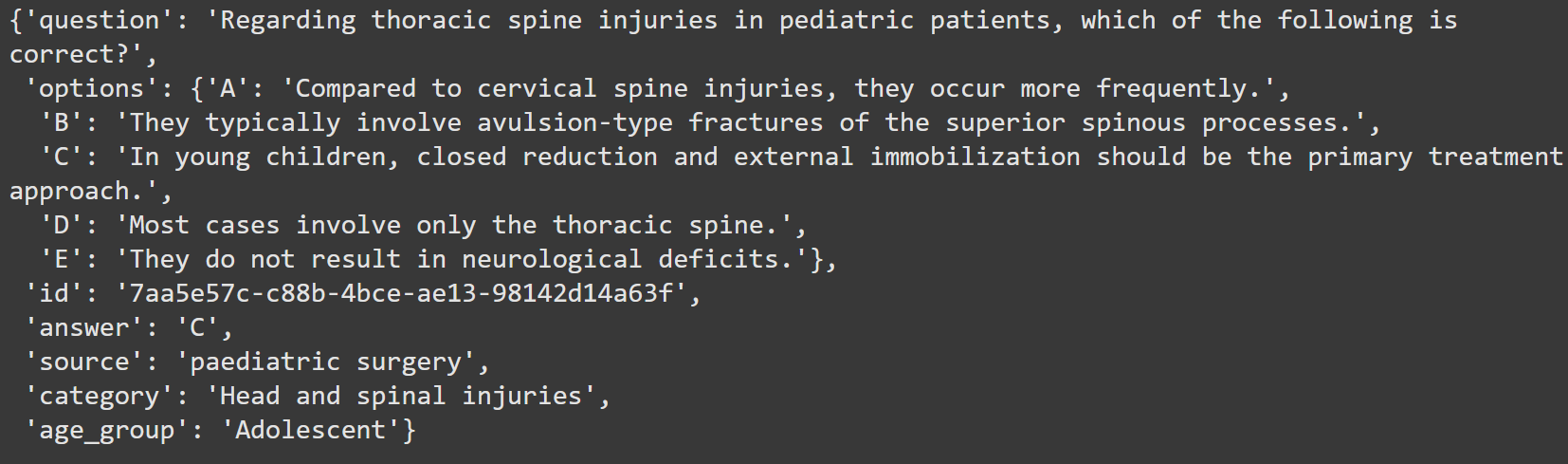}
        \includegraphics[scale=0.4]{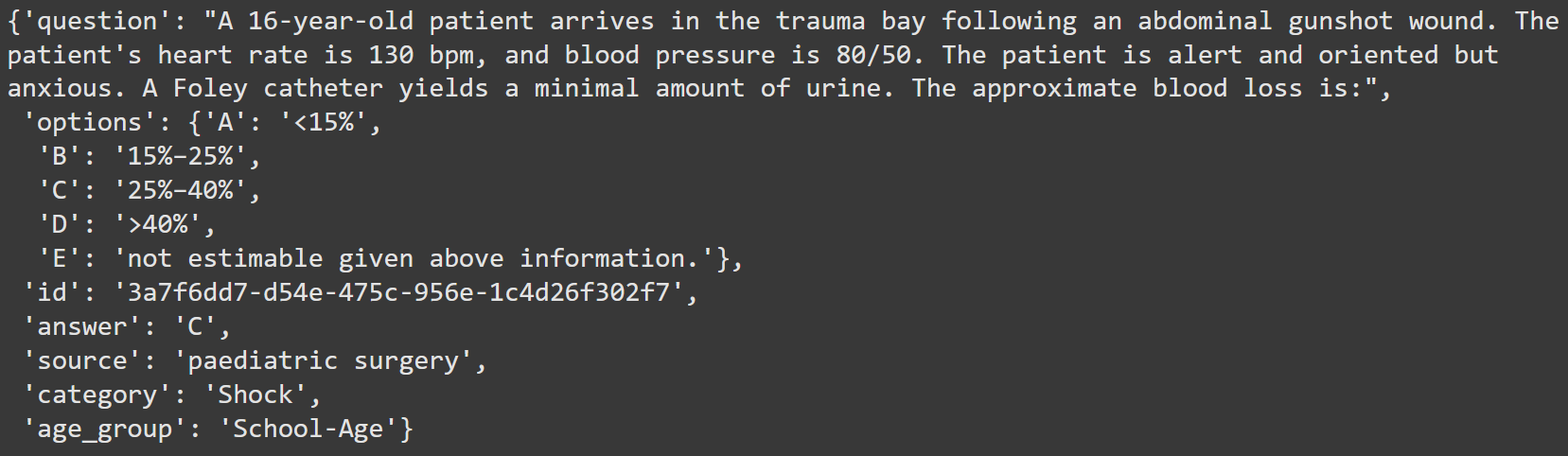}
        \caption{Examples of TQA pairs from the PediatricsMQA dataset.}
        \label{fig:tqa_ex}
    \end{figure}

\begin{figure}
        \centering
        \subfloat[\centering Example 1: Image]{{\includegraphics[scale=0.25]{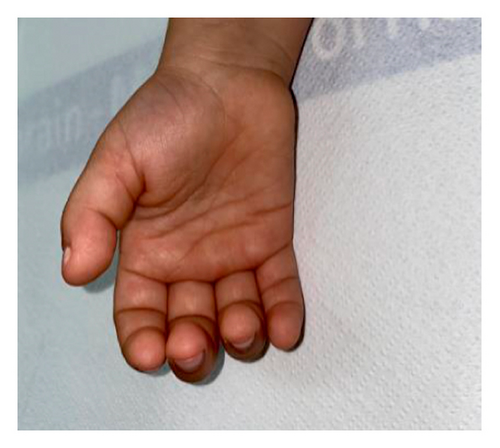} }}%
        \qquad
        \subfloat[\centering Example 1: Metadata]{{\includegraphics[scale=0.3]{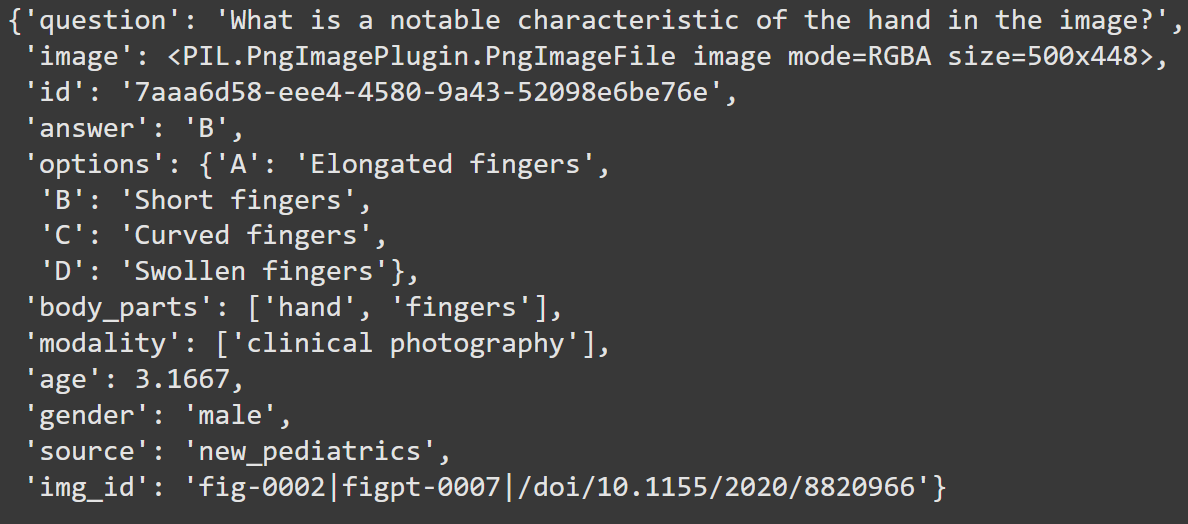} }}%
        \qquad
        \subfloat[\centering Example 2: Image]{{\includegraphics[scale=0.25]{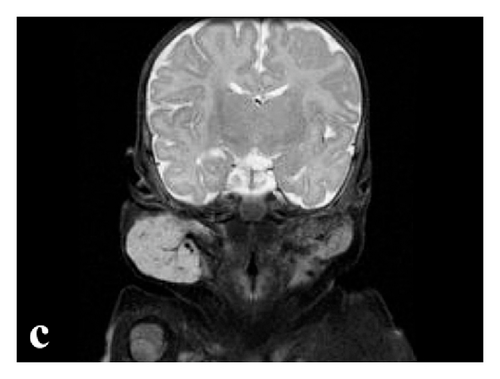} }}%
        \qquad
        \subfloat[\centering Example 2: Metadata]{{\includegraphics[scale=0.25]{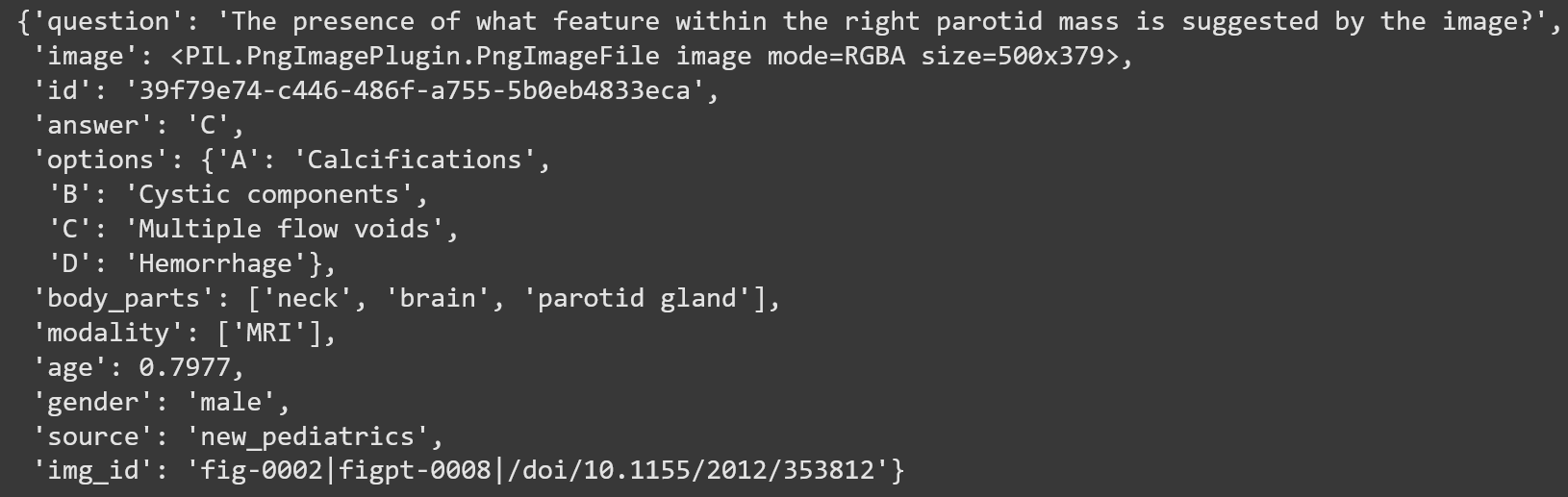} }}%
        \caption{Examples of VQA pairs from the PediatricsMQA dataset.}
        \label{fig:vqa_ex}
    \end{figure}

    The dataset is available on HuggingFace\footnote{\url{https://huggingface.co/datasets/adlbh/PediatricsMQA}}. Figures \ref{fig:tqa_ex} and \ref{fig:vqa_ex} give two examples of TQA and VQA pairs in the data.
\section{Prompts and Templates}
    \label{app:prompts}
\begin{figure}
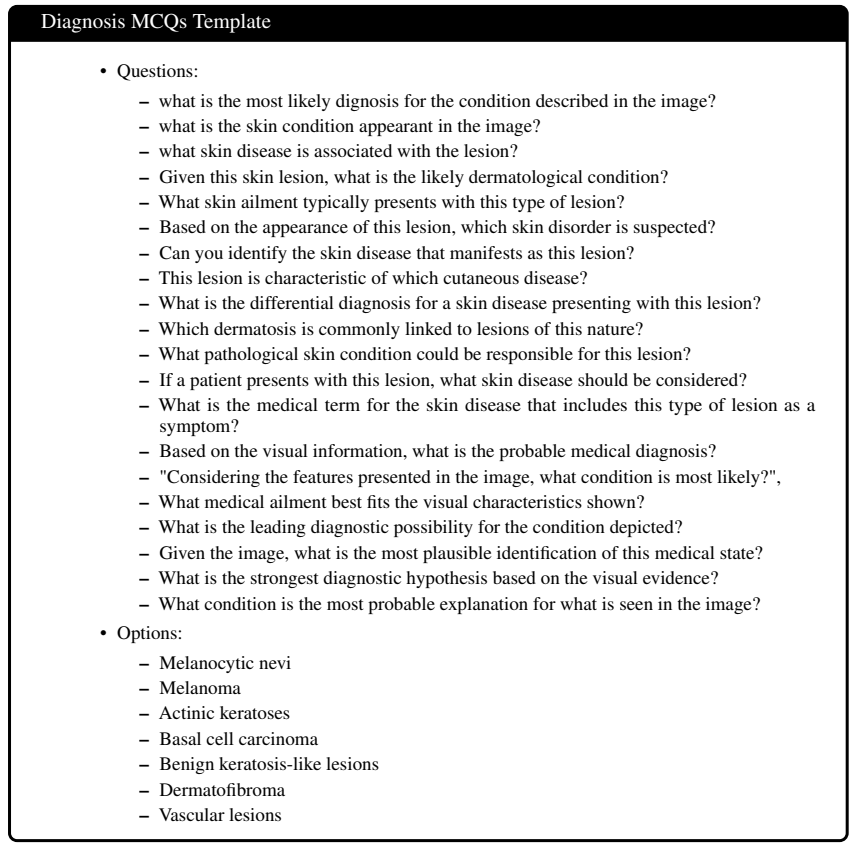

        \centering
        \scalebox{0.8}{%
        \begin{tcolorbox}[colback=white, colframe=black, title=Diagnosis MCQs Template]
        \small
            \begin{itemize}
                \item Questions:
                \begin{itemize}
                    \item what is the most likely dignosis for the condition described in the image?
                    \item what is the skin condition appearant in the image?
                    \item what skin disease is associated with the lesion?
                    \item Given this skin lesion, what is the likely dermatological condition?
                    \item What skin ailment typically presents with this type of lesion?
                    \item Based on the appearance of this lesion, which skin disorder is suspected?
                    \item Can you identify the skin disease that manifests as this lesion?
                    \item This lesion is characteristic of which cutaneous disease?
                    \item What is the differential diagnosis for a skin disease presenting with this lesion?
                    \item Which dermatosis is commonly linked to lesions of this nature?
                    \item What pathological skin condition could be responsible for this lesion?
                    \item If a patient presents with this lesion, what skin disease should be considered?
                    \item What is the medical term for the skin disease that includes this type of lesion as a symptom?
                    \item Based on the visual information, what is the probable medical diagnosis?
                    \item "Considering the features presented in the image, what condition is most likely?",
                    \item What medical ailment best fits the visual characteristics shown?
                    \item What is the leading diagnostic possibility for the condition depicted?
                    \item Given the image, what is the most plausible identification of this medical state?
                    \item What is the strongest diagnostic hypothesis based on the visual evidence?
                    \item What condition is the most probable explanation for what is seen in the image?
                \end{itemize}
                \item Options:
                \begin{itemize}
                    \item Melanocytic nevi
                    \item Melanoma
                    \item Actinic keratoses
                    \item Basal cell carcinoma
                    \item Benign keratosis-like lesions
                    \item Dermatofibroma
                    \item Vascular lesions
                \end{itemize}
        \end{itemize}
        \end{tcolorbox}
        }
        \caption{Questions and options templates for HAM10000 \cite{tschandl2018ham10000} to generate diagnosis MCQs.}
        \label{fig:ham_gen_prompt_diag}
    \end{figure}

    This section lists various prompts (Figures \ref{fig:para_prompt}, \ref{fig:vqa_prompt}, \ref{fig:fairvl_gen_prompt}, \ref{fig:tva_inference_prompt}, \ref{fig:vqa_inference_prompt}) and templates (\ref{fig:ham_gen_prompt_diag}, \ref{fig:ham_gen_prompt_loc}) used during the construction and inference process. The captions of the images are self-descriptive.
    \begin{figure}[htbp]
        \centering
        \scalebox{0.8}{%
        \begin{tcolorbox}[colback=white, colframe=black, title=TQA pairs Paraphrasing Prompt]
        \small
        \#\#\# \textbf{Instruction:} \\
        - Paraphrase the following question and options without changing their meaning or the order of options.\\
        - keep medical terms.\\
        - Your output should have the following format:\\
          \{
            "question": "",
            "options": \{
              "A": "",
              "B": "",
              "C": "",
              "D": "",
              "E": ""
              ...
            \}
          \}
          \\[1mm]
        \#\#\# \textbf{Input:} \\
        \textcolor{blue}{<ORIGINAL Question and Corresponding Options>}\\[1mm]
        \#\#\# Response:
        \end{tcolorbox}
        }
        \caption{Prompt for paraphrasing TQA pairs of PediatricsMQA.}
        \label{fig:para_prompt}
    \end{figure}

    \begin{figure}[htbp]
        \centering
        \scalebox{0.8}{%
        \begin{tcolorbox}[colback=white, colframe=black, title=VQA Pairs Generation Prompt]
        \small
        \#\#\# \textbf{Instruction:} \\
        - You are given an image and its corresponding title, caption and quotes.\\
        - Generate 5 multiple-choice questions from the available information only, and provide your reasoning for each question.\\
        - The questions need to be based on the caption, quotes.\\
        - Only refer to the image in the questions, even if you use the caption or quote to create the question. Meaning your question should never contain any sentence similar or equivalent to "based on the caption".\\
        - The questions should be as challenging as possible.\\
        - Give your confidence (from 0 to 100) in each generated question's validity and corresponding answer.\\
        - Your output should have the following JSON format:\\
        $[$
        \{
          "confidence": "",
          "question":"",
          "options":\{
            "A": "",
            "B": "",
            "C": "",
            "D": ""
          \},
          "answer": "A" or "B" or ...,
          "reasoning: ""
        \},
        ...
        $]$\\
        \textbf{Title:}\\
        \textcolor{blue}{<IMAGE TITLE>}\\
        \textbf{Caption:}\\
        \textcolor{blue}{<IMAGE CAPTION>}\\
        \textbf{Quotes:}\\
        \textcolor{blue}{<IMAGE ARTICLE QUOTES>}\\
        \#\#\# Response:
        \end{tcolorbox}
        }
        \caption{Prompt for Generating VQA pairs of PediatricsMQA based on scientific articles.}
        \label{fig:vqa_prompt}
    \end{figure}
    
    \begin{figure}
        \centering
        \scalebox{0.8}{%
        \begin{tcolorbox}[colback=white, colframe=black, title=FairVLMed VQA pairs generation prompt]
        \small
        - From the image and the context generate 5 multiple choice questions.\\
        - The questions should be visually answerable.\\
        - The answer to the question should exist in the knowledge presented in the context, don't use any external knowledge.\\
        - Non of the questions should be about the existence of glocoma in the image.\\
        - Your output should have the folowing format:\\
        $[$\\
        \{\\
          "question":"",\\
          "options":$[$\\
            "A": "",\\
            "B": "",\\
            "C": "",\\
            "D": ""\\
          $]$,\\
          "answer": "A" or "B" or ...,\\
          "reasoning: ""\\
        \},\\
        ...\\
        $]$\\
        Context:\\
        \textcolor{blue}{<MEDICAL NOTES>}\\
        \end{tcolorbox}
        }
        \caption{Prompt for Generating VQA pairs of PediatricsMQA based on FairVLMed \cite{luo2024fairclip}.}
        \label{fig:fairvl_gen_prompt}
    \end{figure}
    
    \begin{figure}
        \centering
        \scalebox{0.8}{%
        \begin{tcolorbox}[colback=white, colframe=black, title=Localization MCQs Template]
        \small
        \begin{itemize}
                \item Questions:
                \begin{itemize}
                    \item Where on the body is the pigmented lesion visible in this dermatoscopic view located?
                    \item Can you identify the anatomical site of the pigmented lesion in the provided dermatoscopic image?
                    \item On which part of the body is this pigmented lesion found, as seen in the dermatoscopic image?
                    \item What is the precise location on the skin where this pigmented lesion is situated (based on the dermatoscopic image)?
                    \item Referring to the dermatoscopic image, what is the body region exhibiting this pigmented lesion?
                    \item Could you specify the anatomical location of the pigmented lesion depicted in this dermatoscopic image?
                    \item Where on the patient's body is the pigmented lesion captured in this dermatoscopic view?
                    \item What is the regional anatomy of the pigmented lesion presented in this dermatoscopic image?
                    \item Based on the dermatoscopic image, which body surface area displays this pigmented lesion?
                    \item Can you pinpoint the exact body location of the pigmented lesion in this dermatoscopic image?
                    \item What is the dermatoscopic location of the pigmented lesion shown?
                    \item Identify the body site of the pigmented lesion as visualized under dermatoscopy.
                    \item Where is the pigmented skin lesion located according to the dermatoscopic image?
                    \item This dermatoscopic image shows a pigmented lesion on which part of the body?
                    \item What bodily region does the pigmented lesion in this dermatoscopic image occupy?
                    \item Can you describe the anatomical position of the pigmented lesion in the given dermatoscopic view?
                    \item Which area of the skin exhibits the pigmented lesion in the dermatoscopic image?
                    \item What is the topographical location of the pigmented lesion in this dermatoscopic image?
                    \item Where on the human anatomy is the pigmented lesion in the dermatoscopic image situated?
                    \item Please indicate the specific body surface location of the pigmented lesion as seen through the dermatoscope.
                \end{itemize}
                \item Options:
                \begin{itemize}
                    \item Unknown
                    \item Scalp
                    \item Chest
                    \item Acral
                    \item Ear
                    \item Foot
                    \item Neck
                    \item Hand
                    \item Back
                    \item Lower extremity
                    \item Trunk
                    \item Abdomen
                    \item Face
                    \item Upper extremity
                    \item Genital
                \end{itemize}
            \end{itemize}
        \end{tcolorbox}
        }
        \caption{Questions and options templates for HAM10000 \cite{tschandl2018ham10000} to generate localization MCQs.}
        \label{fig:ham_gen_prompt_loc}
    \end{figure}

    \begin{figure}
        \centering
        \scalebox{0.8}{%
        \begin{tcolorbox}[colback=white, colframe=black, title=TQA inference prompt]
        \small
        - Answer the question by choosing the right option index.\\
        - Only give the index letter of the right option.\\
        - Don't show your reasoning, give your answer directly.\\
        
        \# \textbf{Question}:\\
         \textcolor{blue}{<Question>}\\
        \# \textbf{Options}:\\
        \textcolor{blue}{<OPTIONS LISTS WITH THEIR INDEX>}\\
        \# \textbf{Response}:\\
        \end{tcolorbox}
        }
        \caption{Prompt for inference in LLMs for TQA answering.}
        \label{fig:tva_inference_prompt}
    \end{figure}

    \begin{figure}
        \centering
        \scalebox{0.8}{%
        \begin{tcolorbox}[colback=white, colframe=black, title=VQA inference prompt]
        \small
        - Given the image, answer the following question by choosing the right option letter (A, or B or C or D)\\
        - Your answer should be direct.\\
        \# \textbf{Question}:\\
        \textcolor{blue}{<Question>}\\
        \# \textbf{Options}:\\
        \textcolor{blue}{<OPTIONS LISTS WITH THEIR INDEX>}\\
        \# \textbf{Answer}:\\
        \end{tcolorbox}
        }
        \caption{Prompt for inference in VLMs for VQA answering.}
        \label{fig:vqa_inference_prompt}
    \end{figure}
    
\section{Additional Results}
    \subsection{Text QA}
        \subsubsection{Best and worst performance per topic category}
        Tables \ref{tab:tqa_wb_cats_gemini_2.0}, \ref{tab:tqa_wb_cats_gemini_1.5}, \ref{tab:tqa_wb_cats_medalpaca-7b}, \ref{tab:tqa_wb_cats_llama-medx}, \ref{tab:tqa_wb_cats_llama_3.1-8b}, \ref{tab:tqa_wb_cats_llama-4-scout-17b}, and \ref{tab:tqa_wb_cats_llama-4-maverick-17b} represent a list of 10 worst and best performing PediatricqMQA(TQA) categories for Gemini-2.0-Flash, Gemini-1.0-Flash, MedAlpaca-7b, Llama-MedX, Llama-3.1 (8b), Llama-4-scout (17b) and Llama-4-maverick (17b) respectively. The results show that these models can score as low as 0\% in some categories, while performing as high as 100\% in other categories. In addition, models perform consistently badly at some categories like "Lipid Disorders" and "Anorectal disorders", while being consistently good at other categories such as "Developmental Psychology". This shows that there are some medical topics that various LLMs find consistently hard, requiring more training or better reasoning in these areas. These findings highlight both the potential and current limitations of LLMs in specialised pediatric medical reasoning, with implications for targeted dataset enrichment and model fine-tuning. The observed performance disparities across pediatric subdomains likely stem from variations in the complexity, specificity, and data representation of the topics within the models’ training corpora. Categories such as Lipid Disorders, Pharmacology, and Neuroradiology, which consistently appear among the worst-performing, often involve dense biochemical knowledge, imaging interpretation, or nuanced pharmacokinetics—areas that require detailed, structured understanding and are less likely to be covered extensively or uniformly in publicly available medical text data. Conversely, high-performing categories like Developmental Psychology, Porphyria, and Paediatric Orthopaedic Disorders may benefit from more narrative-style content, clearer clinical patterns, or broader representation in educational materials and online resources. Furthermore, the consistent success in Developmental Psychology suggests that topics grounded in developmental milestones and behavioral patterns—often discussed in accessible formats—are more easily internalised by language models. These trends indicate that performance may be correlated not only with domain difficulty but also with the nature and quality of training data available for each category. Targeted augmentation with high-quality domain-specific data could help address these disparities and improve model reliability across underperforming clinical areas.
        \begin{table}[!h]
        \centering
        \setlength{\tabcolsep}{1.7pt}
        \begin{tabular}{rl|rl}
            \toprule
            \multicolumn{2}{c}{Worst} & \multicolumn{2}{|c}{Best} \\
            \hline
            Acc & Category & Acc & Category \\
            \midrule
            0.00 & Lipid Disorders & 100.00 & Neurogenic bladder \\
            33.33 & Alimentary tract duplications & 100.00 & Genetic Syndromes \\
            33.33 & Endocrinology & 100.00 & Porphyria \\
            35.00 & Anorectal disorders & 100.00 & Paediatric orthopaedic disorders \\
            37.50 & Vascular access & 100.00 & Pharmacology \\
            37.50 & Intussusception & 100.00 & Fluids, electrolytes and nutritional support \\
            38.46 & Gallbladder disease & 100.00 & Neuroradiology \\
            40.00 & Appendicitis & 94.12 & Thoracic trauma \\
            41.18 & Intestinal atresia & 93.33 & Disorders of sexual differentiation \\
            42.11 & Anorectal continence and constipation & 92.86 & musculoskeletal trauma and soft tissue injuries \\
            \bottomrule
        \end{tabular}
        \caption{Best and worst performing topic categories of PediatricsMQA(TQA) on Gemini-2.0-flash.}
        \label{tab:tqa_wb_cats_gemini_2.0}
    \end{table}

    \begin{table}[!h]
        \centering
        \setlength{\tabcolsep}{1.7pt}
        \begin{tabular}{rlrl}
            \toprule
            \multicolumn{2}{r}{Worst} & \multicolumn{2}{r}{Best} \\
            Acc & Category & Acc & Category \\
            \midrule
            0.00 & Lipid Disorders & 100.00 & Hypospadias \\
            0.00 & Neuroradiology & 100.00 & Porphyria \\
            0.00 & Genetic Syndromes & 100.00 & Developmental Psychology|Adolescence \\
            0.00 & Pharmacology & 92.86 & Paediatric orthopaedic disorders \\
            6.25 & Vascular access & 88.89 & Neurogenic bladder \\
            12.50 & Intussusception & 88.24 & Thoracic trauma \\
            22.22 & Congenital diaphragmatic hernia & 85.71 & The Field of Pediatrics \\
            25.00 & malrotation and midgut volvulus & 85.00 & Developmental Psychology|School Child \\
            25.00 & Anorectal disorders & 84.62 & Coagulopathies and surgical infectious diseases \\
            28.57 & mechanical ventilation and support & 83.33 & Chest wall deformities \\
            \bottomrule
        \end{tabular}
        \caption{Best and worst performing topic categories of PediatricsMQA(TQA) on Gemini-1.5-flash.}
        \label{tab:tqa_wb_cats_gemini_1.5}
    \end{table}

    \begin{table}[!h]
        \centering
        \begin{tabular}{rlrl}
            \toprule
            \multicolumn{2}{r}{Worst} & \multicolumn{2}{r}{Best} \\
            Acc & Category & Acc & Category \\
            \midrule
            0.00 & Genetic Syndromes & 75.00 & Posterior urethral valves \\
            0.00 & Porphyria & 75.00 & Developmental Psychology|Adolescence \\
            0.00 & Neuroradiology & 72.73 & Circumcision and disorders of penis \\
            0.00 & Lipid Disorders & 63.16 & Developmental Psychology|Pre-School Child \\
            0.00 & Pharmacology & 61.54 & Gastro- oesophageal reflux disease \\
            6.67 & Bariatric surgery in children & 60.00 & Cardiovascular system \\
            7.69 & Gallbladder disease & 60.00 & Developmental Psychology|School Child \\
            10.00 & Appendicitis & 57.14 & Caustic injuries of the oesophagus \\
            11.11 & Rehabilitation Medicine & 56.25 & Renal diseases in children \\
            11.76 & General approach to trauma & 55.00 & Developmental Psychology|Infancy 0-2 Years \\
            \bottomrule
        \end{tabular}
        \caption{Best and worst performing topic categories of PediatricsMQA(TQA) on Medalpaca-7b.}
        \label{tab:tqa_wb_cats_medalpaca-7b}
    \end{table}
    
    \begin{table}[!h]
        \centering
        \setlength{\tabcolsep}{1.7pt}
        \scalebox{0.8}{
        \begin{tabular}{rlrl}
            \toprule
            \multicolumn{2}{r}{Worst} & \multicolumn{2}{r}{Best} \\
            Acc & Category & Acc & Category \\
            \midrule
            0.00 & Pharmacology & 100.00 & Porphyria \\
            0.00 & Lipid Disorders & 100.00 & Genetic Syndromes \\
            0.00 & Neuroradiology & 92.86 & Paediatric orthopaedic disorders \\
            10.00 & Hernia and hydrocele & 80.00 & Hypospadias \\
            12.50 & Intussusception & 78.95 & Developmental Psychology|Pre-School Child \\
            18.75 & Vascular access & 78.57 & urinary incontinence \\
            22.22 & Congenital diaphragmatic hernia & 77.78 & Rehabilitation Medicine \\
            25.00 & Disorders of colonic motility & 76.47 & Gastrointestinal bleeding \\
            25.00 & Gynecologic Problems of Childhood & 75.00 & Developmental Psychology|Infancy 0-2 Years \\
            25.00 & Developmental and positional anomalies of the kidney & 75.00 & Developmental Psychology|Adolescence \\
            \bottomrule
        \end{tabular}}
        \caption{Best and worst performing topic categories of PediatricsMQA(TQA) on Llama-medx.}
        \label{tab:tqa_wb_cats_llama-medx}
    \end{table}
    \begin{table}[!h]
        \centering
        \setlength{\tabcolsep}{1.7pt}
        \begin{tabular}{rlrl}
        \toprule
        \multicolumn{2}{r}{Worst} & \multicolumn{2}{r}{Best} \\
        Acc & Category & Acc & Category \\
        \midrule
        0.00 & lesions of the stomach & 85.00 & Developmental Psychology|Adolescence \\
        0.00 & Neuroradiology & 78.95 & Developmental Psychology|Pre-School Child \\
        0.00 & Genetic Syndromes & 72.73 & Circumcision and disorders of penis \\
        0.00 & Pharmacology & 70.00 & Developmental Psychology|School Child \\
        0.00 & Lipid Disorders & 65.00 & Developmental Psychology|Infancy 0-2 Years \\
        6.25 & Vascular access & 64.29 & Necrotising enterocolitis \\
        10.00 & Hernia and hydrocele & 64.29 & urinary incontinence \\
        10.53 & Anorectal disorders & 60.00 & Hypospadias \\
        12.50 & Congenital abdominal wall defects & 60.00 & Alimentary tract duplications \\
        13.33 & Bariatric surgery in children & 60.00 & Developmental Psychology \\
        \bottomrule
        \end{tabular}
        \caption{Best and worst performing topic categories of PediatricsMQA(TQA) on Llama-3.1-8b.}
        \label{tab:tqa_wb_cats_llama_3.1-8b}
    \end{table}
    \begin{table}[!h]
        \centering
        \setlength{\tabcolsep}{1.7pt}
        \scalebox{0.8}{
        \begin{tabular}{rlrl}
        \toprule
        \multicolumn{2}{r}{Worst} & \multicolumn{2}{r}{Best} \\
        Acc & Category & Acc & Category \\
        \midrule
        0.00 & Lipid Disorders & 100.00 & Genetic Syndromes \\
        12.50 & Intussusception & 100.00 & Paediatric orthopaedic disorders \\
        25.00 & Vascular access & 100.00 & Pharmacology \\
        30.00 & Growth and development & 100.00 & Thoracic trauma \\
        33.33 & Testicular problems and varicoceles & 100.00 & Neuroradiology \\
        33.33 & Endocrinology & 100.00 & Porphyria \\
        33.33 & Adjuvant therapy in childhood cancer & 92.86 & musculoskeletal trauma and soft tissue injuries \\
        33.33 & Tracheobronchial and oesophageal foreign bodies & 90.91 & Circumcision and disorders of penis \\
        35.29 & Hepatic tumours & 90.00 & Developmental Psychology|School Child \\
        36.84 & Anorectal continence and constipation & 90.00 & Paediatric neurosurgical disorders \\
        \bottomrule
        \end{tabular}}
         \caption{Best and worst performing topic categories of PediatricsMQA(TQA) on Llama-4-scout-17b.}
        \label{tab:tqa_wb_cats_llama-4-scout-17b}
    \end{table}
    
    \begin{table}[!h]
        \centering
        \setlength{\tabcolsep}{1.7pt}
        \begin{tabular}{rlrl}
        \toprule
        \multicolumn{2}{r}{Worst} & \multicolumn{2}{r}{Best} \\
        Acc & Category & Acc & Category \\
        \midrule
        0.00 & Pharmacology & 100.00 & General approach to trauma \\
        25.00 & Vascular access & 100.00 & Lipid Disorders \\
        40.00 & Neonatology & 100.00 & Neurogenic bladder \\
        40.00 & Vascular disorders & 100.00 & Rehabilitation Medicine \\
        40.00 & Cardiovascular system & 100.00 & Genetic Syndromes \\
        41.18 & Intestinal atresia & 100.00 & Porphyria \\
        44.44 & Congenital diaphragmatic hernia & 100.00 & Paediatric orthopaedic disorders \\
        45.00 & Anorectal disorders & 100.00 & Neuroradiology \\
        45.83 & Gynecologic Problems of Childhood & 95.00 & Endocrine disorders \\
        46.15 & Gallbladder disease & 94.74 & Fetal surgery \\
        \bottomrule
        \end{tabular}
        \caption{Best and worst performing topic categories of PediatricsMQA(TQA) on Llama-4-maverick-17b.}
        \label{tab:tqa_wb_cats_llama-4-maverick-17b}
    \end{table}
    
    \subsubsection{Performance by age group}
        Table \ref{tab:age_group_tqa} illustrates the performance of various large language models on a pediatrics question answering dataset, with results broken down by different pediatric age groups: Prenate, Neonate, Infant, Toddler, Preschool, School-Age, and Adolescent. A consistent trend across all models is that performance varies between age groups, with no single age group being universally easier or harder for all models. Generally, the Llama-4-maverick (17b) and Gemini-2.0-flash models demonstrate the highest performance across most age categories, often achieving scores above 70\%. Conversely, MedAlpaca (7b) and Llama-3.1 (8b) consistently show lower performance, typically below 45\% across all age groups. For most models, there isn't a clear linear trend of increasing or decreasing performance with age; for instance, Llama-medx performs well in the Prenate and Adolescent groups but sees a dip in the Toddler category. Similarly, Gemini-1.5-flash shows relatively stable performance across most age groups with slight variations. The larger models, Llama-4-scout (17b) and Llama-4-maverick (17b), along with the Gemini series, generally outperform the smaller models like MedAlpaca (7b) and Llama-3.1 (8b), indicating a correlation between model size/sophistication and performance on this specialised medical question-answering task.   
            \begin{table}[htbp]
                \centering
                \setlength{\tabcolsep}{1.7pt}
               \begin{tabular}{lrrrrrrr}
                \toprule
                 & Prenate & Neonate & Infant & Toddler & Preschool & School-Age & Adolescent \\
                \midrule
                MedAlpaca (7b) & 28.74 & 33.56 & 31.97 & 33.08 & 36.17 & 29.44 & 29.18 \\
                Llama-medx & 50.00 & 48.29 & 50.21 & 45.86 & 56.74 & 48.77 & 54.71 \\
                Gemini-1.5-flash & 60.54 & 54.11 & 55.29 & 54.51 & 58.16 & 58.15 & 59.42 \\
                Gemini-2.0-flash & 73.75 & 65.13 & 69.87 & 70.68 & 70.92 & 70.71 & 74.67 \\
                Llama-3.1 (8b) & 37.93 & 38.08 & 35.12 & 41.35 & 41.13 & 38.10 & 40.98 \\
                Llama-4-scout (17b) & 72.80 & 63.01 & 64.08 & 57.89 & 66.67 & 66.96 & 70.74 \\
                Llama-4-maverick (17b) & 77.22 & 68.49 & 69.96 & 66.17 & 70.92 & 75.04 & 77.98 \\
                \bottomrule
                \end{tabular}
                \caption{The performance of different models for different age groups on the PediatricsMQA (TQA).}
                \label{tab:age_group_tqa}
            \end{table}
    \subsection{Vision QA}
        \subsubsection{Performance by age group}
            Table \ref{tab:vqa_age_group_res} presents the performance of various vision-language models on the PediatricsMQA (Visual Question Answering) dataset, disaggregated by pediatric age groups: Prenate, Neonate, Infant, Toddler, Preschool, School-Age, and Adolescent. Among the models evaluated, Huatuogpt-vision (7B) consistently achieves the highest accuracy across most age groups, notably peaking at 73.77\% in the Neonate and 73.48\% in the Toddler categories. Gemini-2.0-flash also demonstrates relatively strong and stable performance, with scores ranging from 47.24\% to 68.18\%, outperforming its predecessor, Gemini-1.5-flash, across all age groups. In contrast, Llava-med-v1.5-mistral (7B) shows the lowest overall accuracy and more variable results, with its best performance (50.45\%) in the Infant category and a marked decline in Preschool (26.76\%). A general trend indicates that model performance tends to vary notably by age group, with the Neonate and Infant groups typically yielding higher accuracies, while performance tends to drop in the Adolescent and Preschool categories, highlighting the challenges models face in generalizing across the developmental spectrum.
            \begin{table}[htbp]
                \centering
                \setlength{\tabcolsep}{1.7pt}
                \begin{tabular}{lrrrrrrr}
                \toprule
                 & Prenate & Neonate & Infant & Toddler & Preschool & School-Age & Adolescent \\
                \midrule
                Gemini-2.0-flash & 56.55 & 68.18 & 61.25 & 57.83 & 54.02 & 58.44 & 47.24 \\
                Gemini-1.5-flash & 23.81 & 40.00 & 31.43 & 16.67 & 34.69 & 33.33 & 32.89 \\
                Huatuogpt-vision (7B) & 59.59 & 73.77 & 68.10 & 73.48 & 54.68 & 61.78 & 54.04 \\
                Llava-med-v1.5-mistral (7b) & 29.94 & 32.35 & 50.45 & 42.40 & 26.76 & 35.87 & 35.28 \\
                \bottomrule
                \end{tabular}
                \caption{Results per age group on the PediatricsMQA(VQA) dataset for different models.}
                \label{tab:vqa_age_group_res}
            \end{table}
        \subsubsection{Performance by modality}
                The tables \ref{tab:vqa_bw_mod_gemini2.0}, \ref{tab:vqa_bw_mod_gemini1.5}, \ref{tab:vqa_bw_mod_huatuogpt7b}, and \ref{tab:vqa_bw_mod_llavav1.5} summarize the performance of various image modalities on the PediatricsMQA(VQA) dataset across four vision-language models: Gemini-2.0-flash, Gemini-1.5-flash, Huatuogpt-vision (7B), and Llava-med-v1.5-mistral (7B), respectively. Each table lists the ten best- and worst-performing modalities based on accuracy. A consistent trend across all models is the stark contrast in performance between modalities associated with visually rich, structured, or clinically interpretable data (e.g., optical, physical exam, surgical specimen, ihc staining)—which frequently achieve 100\% accuracy—and those involving abstract, low-contrast, or complex image types (e.g., natural image, cytopathology, vce, sanger sequencing)—which often yield 0\% accuracy. Additionally, modalities like radiograph, angiogram, and echocardiography appear among both the best and worst categories depending on the model, suggesting that model-specific strengths and weaknesses influence modality interpretation. This disparity highlights significant variability in how different vision-language architectures process and reason over medical imagery.

            \begin{table}[htbp]
                \centering
                \begin{tabular}{rlrl}
                \toprule
                \multicolumn{2}{r}{Worst} & \multicolumn{2}{r}{Best} \\
                acc & modality & acc & modality \\
                \midrule
                0.00 & vce & 100.00 & optical \\
                0.00 & natural image & 100.00 & physical exam \\
                0.00 & cytopathology & 100.00 & sanger sequencing \\
                22.22 & radiograph & 100.00 & biopsy \\
                25.00 & ct angiogram & 100.00 & curve \\
                30.00 & echocardiography & 100.00 & surgical photography \\
                40.00 & angiogram & 100.00 & ihc staining \\
                40.70 & fundus photography & 100.00 & bone scan \\
                47.02 & clinical photography & 100.00 & cta \\
                50.00 & doppler & 100.00 & ultrasonography \\
                \bottomrule
                \end{tabular}
                \caption{Ten best and worst performing modalities of PediatricsMQA(VQA) on Gemini-2.0-flash.}
                \label{tab:vqa_bw_mod_gemini2.0}
            \end{table}

            \begin{table}[htbp]
                \centering
                \begin{tabular}{rlrl}
                \toprule
                \multicolumn{2}{r}{Worst} & \multicolumn{2}{r}{Best} \\
                acc & modality & acc & modality \\
                \midrule
                0.00 & angiography & 100.00 & doppler \\
                0.00 & diagram & 100.00 & surgical specimen \\
                0.00 & illustration & 100.00 & photo \\
                0.00 & ultrasound & 100.00 & fish \\
                0.00 & surgical photograph & 75.00 & curve \\
                0.00 & mra & 60.00 & microscopy \\
                0.00 & cytopathology & 60.00 & ecg \\
                0.00 & sonography & 60.00 & photograph \\
                0.00 & biopsy & 50.00 & echocardiogram \\
                0.00 & ct angiography & 50.00 & he stain \\
                \bottomrule
                \end{tabular}
                \caption{Ten best and worst performing modalities of PediatricsMQA(VQA) on Gemini-1.5-flash.}
                \label{tab:vqa_bw_mod_gemini1.5}
            \end{table}

            \begin{table}[htbp]
                \centering
                \begin{tabular}{rlrl}
                \toprule
                \multicolumn{2}{r}{Worst} & \multicolumn{2}{r}{Best} \\
                acc & modality & acc & modality \\
                \midrule
                0.00 & sanger sequencing & 100.00 & he stain \\
                0.00 & natural image & 100.00 & ihc staining \\
                0.00 & curve & 100.00 & cytopathology \\
                25.00 & still image & 100.00 & surgical specimen \\
                33.33 & vce & 100.00 & electrocardiogram \\
                40.00 & bone scan & 100.00 & ultrasonography \\
                42.86 & illustration & 100.00 & mra \\
                45.45 & radiograph & 100.00 & ct-angiography \\
                48.54 & clinical photography & 100.00 & cta \\
                50.00 & dna sequencing & 100.00 & pet \\
                \bottomrule
                \end{tabular}
                \caption{Ten best and worst performing modalities of PediatricsMQA(VQA) on Huatuogpt-vision (7B).}
                \label{tab:vqa_bw_mod_huatuogpt7b}
            \end{table}

            \begin{table}[htbp]
                \centering
                \begin{tabular}{rlrl}
                \toprule
                \multicolumn{2}{r}{Worst} & \multicolumn{2}{r}{Best} \\
                acc & modality & acc & modality \\
                \midrule
                0.00 & radiography & 100.00 & fluoroscopy \\
                0.00 & pet & 100.00 & dexa \\
                0.00 & color doppler & 100.00 & cytopathology \\
                0.00 & colonoscopy & 100.00 & still image \\
                0.00 & natural image & 100.00 & optical \\
                0.00 & ct angiogram & 100.00 & physical exam \\
                0.00 & roentgenogram & 100.00 & he stain \\
                0.00 & ultrastructural examination & 100.00 & mra \\
                0.00 & electrocardiogram & 75.00 & angiogram \\
                0.00 & sanger sequencing & 75.00 & radiograph \\
                \bottomrule
                \end{tabular}
                \caption{Ten best and worst performing modalities of PediatricsMQA(VQA) on Llava-med-v1.5-mistral (7b)}
                \label{tab:vqa_bw_mod_llavav1.5}
            \end{table}

        \subsubsection{Performance by anatomical regions}
            The tables \ref{tab:vqa_bw_ana_gemini2.0}, \ref{tab:vqa_bw_ana_gemini1.5}, \ref{tab:vqa_bw_ana_huatuogpt_7b}, and \ref{tab:vqa_bw_ana_llava_v1.5_7b} present the best and worst performing anatomical regions in the PediatricsMQA(VQA) dataset across four different vision-language models: Gemini-2.0-flash, Gemini-1.5-flash, Huatuogpt-vision (7B), and Llava-med-v1.5-mistral (7B), respectively. Each table lists anatomical regions sorted by their question answering accuracy, with many regions consistently achieving perfect (1.00) or zero (0.00) accuracy. A cross-model comparison reveals several persistent trends. Notably, regions associated with cardiovascular and lymphatic structures—such as the coronary artery, vertebral body, blood cells, and buccal mucosa cells—frequently appear among the best performing across all models. Conversely, anatomical sites such as the gums, genital regions, parietal and occipital regions, axilla, and bronchus consistently fall among the worst-performing. These patterns suggest that models are more proficient in recognising and reasoning over medically salient or frequently imaged internal structures, whereas more peripheral, less frequently annotated, or visually ambiguous areas tend to challenge model performance uniformly.
        \begin{table}[htbp]
            \centering
            \begin{tabular}{rlrl}
            \toprule
            \multicolumn{2}{r}{Worst} & \multicolumn{2}{r}{Best} \\
            acc & anatomical region & acc & anatomical region \\
            \midrule
            0.00 & bronchus & 1.00 & hip joint \\
            0.00 & breast & 1.00 & polyp \\
            0.00 & torso & 1.00 & rectal cul de sac \\
            0.00 & testis & 1.00 & coronary artery \\
            0.00 & epididymis & 1.00 & upper lip \\
            0.00 & muscles & 1.00 & hemithorax \\
            0.00 & genital & 1.00 & buccal mucosa cells \\
            0.00 & occipital region & 1.00 & lumbar spine \\
            0.00 & gums & 1.00 & l5 \\
            0.00 & frontal region & 1.00 & l4 \\
            0.00 & armpit & 1.00 & common femoral vein \\
            0.00 & brainstem & 1.00 & external iliac vein \\
            0.00 & arms & 1.00 & renal vein \\
            0.00 & jaw & 1.00 & lumbar veins \\
            0.00 & axilla & 1.00 & left coronary artery \\
            0.00 & intestines & 1.00 & right coronary artery \\
            0.00 & parietal region & 1.00 & vessel \\
            0.00 & lmca & 1.00 & coronary arteries \\
            0.25 & umbilical cord & 1.00 & frontal sinuses \\
            0.25 & c5-c6 & 1.00 & ethmoid sinuses \\
            0.25 & prostatic utricle & 1.00 & palm \\
            0.25 & knee & 1.00 & forehead \\
            0.25 & ulna & 1.00 & maxillary sinus \\
            0.25 & ovary & 1.00 & salivary gland \\
            0.25 & t1 & 1.00 & chromosome \\
            0.25 & c7 & 1.00 & oral cavity \\
            0.27 & ear & 1.00 & toe \\
            0.33 & spinal canal & 1.00 & palatal bone \\
            0.33 & omentum & 1.00 & carotid artery \\
            0.33 & umbilical ring & 1.00 & lumbar canal \\
            0.33 & chromosomes & 1.00 & filum terminale \\
            0.33 & bronchi & 1.00 & artery \\
            0.33 & inguinal region & 1.00 & cerebrum \\
            0.33 & leg & 1.00 & femoral nerve \\
            0.33 & hippocampal formation & 1.00 & vertebral body \\
            0.33 & upper limbs & 1.00 & blood cells \\
            0.33 & spiracles & 1.00 & sural nerve \\
            0.33 & pulmonary veins & 1.00 & acetabulum \\
            0.33 & allocortex & 1.00 & rectosigmoid colon \\
            0.33 & placenta & 1.00 & bone marrow \\
            \bottomrule
            \end{tabular}
            \caption{Forty best and worst performing anatomical regions of PediatricsMQA(VQA) on Gemini-2.0-flash.}
            \label{tab:vqa_bw_ana_gemini2.0}
        \end{table}
        
        \begin{table}[htbp]
            \centering
            \begin{tabular}{rlrl}
            \toprule
            \multicolumn{2}{r}{Worst} & \multicolumn{2}{r}{Best} \\
            acc & anatomical region & acc & anatomical region \\
            \midrule
            0.00 & hand & 1.00 & l5 \\
            0.00 & fingers & 1.00 & buccal mucosa cells \\
            0.00 & parotid gland & 1.00 & blood cells \\
            0.00 & skull & 1.00 & pulmonary veins \\
            0.00 & vertebral artery & 1.00 & lymph node \\
            0.00 & scalp & 1.00 & inferior vena cava \\
            0.00 & aortic arch & 1.00 & capsule \\
            0.00 & lv apex & 1.00 & adrenal gland \\
            0.00 & cerebellum & 1.00 & unknown \\
            0.00 & mouth & 1.00 & l4 \\
            0.00 & teeth & 1.00 & vertebral body \\
            0.00 & superior mesenteric vein & 0.67 & trachea \\
            0.00 & placenta & 0.67 & left ventricle \\
            0.00 & bladder & 0.60 & foot \\
            0.00 & thyroid & 0.60 & tongue \\
            0.00 & pulmonary vein & 0.50 & hyoid bone \\
            0.00 & palate & 0.50 & upper extremity \\
            0.00 & uterus & 0.50 & trunk \\
            0.00 & lips & 0.50 & sural nerve \\
            0.00 & muscles & 0.50 & penis \\
            0.00 & stomach & 0.50 & temporal lobes \\
            0.00 & vertebrae & 0.50 & sigmoid colon \\
            0.00 & spleen & 0.50 & cerebral artery \\
            0.00 & ribs & 0.50 & femoral nerve \\
            0.00 & pancreas & 0.50 & thigh \\
            0.00 & arm & 0.50 & perineum \\
            0.00 & right coronary artery & 0.50 & left atrium \\
            0.00 & left coronary artery & 0.50 & sinuses \\
            0.00 & coronary arteries & 0.50 & upper limbs \\
            0.00 & duodenum & 0.50 & facial structures \\
            0.00 & umbilical ring & 0.50 & cerebral peduncles \\
            0.00 & omentum & 0.50 & pons \\
            0.00 & small bowel & 0.50 & cystic spaces \\
            0.00 & lungs & 0.50 & lower limbs \\
            0.00 & ureter & 0.50 & anus \\
            0.00 & nasopharynx & 0.50 & tricuspid valve \\
            0.00 & maxilla & 0.45 & lower extremity \\
            0.00 & cells & 0.43 & liver \\
            0.00 & axilla & 0.41 & heart \\
            0.00 & left atrial appendage & 0.40 & colon \\
            \bottomrule
            \end{tabular}
            \caption{Forty best and worst performing anatomical regions of PediatricsMQA(VQA) on Gemini-1.5-flash.}
            \label{tab:vqa_bw_ana_gemini1.5}
        \end{table}
        
        \begin{table}[htbp]
            \centering
            \begin{tabular}{rlrl}
            \toprule
            \multicolumn{2}{r}{Worst} & \multicolumn{2}{r}{Best} \\
            acc & anatomical region & acc & anatomical region \\
            \midrule
            0.00 & carotid artery & 1.00 & blood cells \\
            0.00 & frontal region & 1.00 & buccal mucosa cells \\
            0.00 & armpit & 1.00 & lumbar spine \\
            0.00 & occipital region & 1.00 & cerebrum \\
            0.00 & parietal region & 1.00 & vertebral body \\
            0.00 & genital & 1.00 & l4 \\
            0.00 & arms & 1.00 & l5 \\
            0.00 & unknown & 1.00 & adrenal gland \\
            0.00 & gums & 1.00 & zygomatic bone \\
            0.00 & jaw & 1.00 & coronary artery \\
            0.20 & sural nerve & 1.00 & upper lip \\
            0.20 & lv apex & 1.00 & rectal cul de sac \\
            0.20 & femoral nerve & 1.00 & cystic spaces \\
            0.25 & shin & 1.00 & clavicles \\
            0.33 & lymph node & 1.00 & brainstem \\
            0.33 & bronchus & 1.00 & whole body \\
            0.33 & torso & 1.00 & incisors \\
            0.33 & knee & 1.00 & intestines \\
            0.33 & left main coronary artery & 1.00 & blood \\
            0.33 & upper limbs & 1.00 & rachis \\
            0.33 & palatal bone & 1.00 & capsule \\
            0.40 & hip joint & 1.00 & vertebra \\
            0.40 & superior mesenteric vein & 1.00 & palm \\
            0.40 & chromosomes & 1.00 & gallbladder \\
            0.40 & ulna & 1.00 & vocal cords \\
            0.40 & muscles & 1.00 & bone marrow \\
            0.40 & forehead & 1.00 & testicle \\
            0.40 & prostatic utricle & 1.00 & right renal artery \\
            0.40 & hyoid bone & 1.00 & interventricular septum \\
            0.40 & hair & 1.00 & lumbar veins \\
            0.40 & transverse colon & 1.00 & external iliac vein \\
            0.40 & jejunum & 1.00 & artery \\
            0.40 & axilla & 1.00 & common femoral vein \\
            0.40 & skeleton & 0.91 & inferior vena cava \\
            0.40 & descending colon & 0.90 & thyroid \\
            0.44 & lower extremity & 0.90 & femoral head \\
            0.44 & placenta & 0.86 & facial structures \\
            0.45 & upper extremity & 0.84 & mediastinum \\
            0.45 & ear & 0.82 & kidneys \\
            0.46 & fingers & 0.82 & mouth \\
            \bottomrule
            \end{tabular}
            \caption{Forty best and worst performing anatomical regions of PediatricsMQA(VQA) on Huatuogpt-vision (7B).}
            \label{tab:vqa_bw_ana_huatuogpt_7b}
        \end{table}
        
        \begin{table}[htbp]
            \centering
            \begin{tabular}{rlrl}
            \toprule
            \multicolumn{2}{r}{Worst} & \multicolumn{2}{r}{Best} \\
            acc & anatomical region & acc & anatomical region \\
            \midrule
            0.00 & carotid artery & 1.00 & cystic spaces \\
            0.00 & artery & 1.00 & fibula \\
            0.00 & spinal canal & 1.00 & upper lip \\
            0.00 & vertebral artery & 1.00 & blood cells \\
            0.00 & aortic arch & 1.00 & rectal cul de sac \\
            0.00 & lower extremities & 1.00 & intestine \\
            0.00 & feet & 1.00 & coronary artery \\
            0.00 & testis & 1.00 & bowel loops \\
            0.00 & lumbar veins & 1.00 & buccal mucosa cells \\
            0.00 & bronchus & 1.00 & lumbar spine \\
            0.00 & epididymis & 1.00 & shin \\
            0.00 & inguinal canal & 1.00 & lmca \\
            0.00 & ureter & 1.00 & rectosigmoid colon \\
            0.00 & bones & 1.00 & axilla \\
            0.00 & nodules & 1.00 & vertebra \\
            0.00 & pulmonary & 1.00 & palm \\
            0.00 & torso & 1.00 & blood \\
            0.00 & common femoral vein & 1.00 & right arm \\
            0.00 & external iliac vein & 1.00 & right pulmonary artery \\
            0.00 & peritoneum & 1.00 & superior vena cava \\
            0.00 & lower limbs & 1.00 & uterus \\
            0.00 & whole body & 1.00 & cervix \\
            0.00 & clavicles & 1.00 & oral cavity \\
            0.00 & testicle & 1.00 & soft tissue \\
            0.00 & arms & 1.00 & bone \\
            0.00 & prostatic utricle & 1.00 & palatal bone \\
            0.00 & buttock & 1.00 & ovary \\
            0.00 & glenoid labrum & 1.00 & forearm \\
            0.00 & epiphysis & 1.00 & toe \\
            0.00 & genital & 1.00 & vaginal canal \\
            0.00 & unknown & 1.00 & omentum \\
            0.00 & hyoid bone & 1.00 & umbilical ring \\
            0.00 & upper limbs & 1.00 & spiracles \\
            0.00 & gums & 1.00 & superior vena cava (svc) \\
            0.00 & ileum & 1.00 & common carotid artery (cc) \\
            0.00 & parietal region & 1.00 & left atrium (la) \\
            0.00 & occipital region & 1.00 & pleura \\
            0.00 & armpit & 1.00 & intestines \\
            0.00 & gastric mucosa & 1.00 & jaw \\
            0.00 & blood vessel & 1.00 & gallbladder \\
            \bottomrule
            \end{tabular}
            \caption{Forty best and worst performing anatomical regions of PediatricsMQA(VQA) on Llava-med-v1.5-mistral (7b).}
            \label{tab:vqa_bw_ana_llava_v1.5_7b}
        \end{table}

\end{document}